\documentclass[aps,10pt,superscriptaddress,prc,eprint,nofootinbib,showkeys]{revtex4-2}
\pdfoutput=1
\usepackage{graphicx}
\usepackage{caption}
\usepackage{subcaption}
\usepackage{ragged2e}
\usepackage{dcolumn}
\usepackage{bm}

\usepackage{amsmath,amsfonts,amsthm,amssymb}
\usepackage{graphics}
\usepackage{cancel}
\usepackage{multirow}
\usepackage{color}
\usepackage[normalem]{ulem} 
\usepackage{braket}

\usepackage{amssymb}

\usepackage{graphicx}
\usepackage{amsmath}
\usepackage{bbold}
\usepackage{yfonts}
\usepackage[colorlinks=true,linktocpage=true,linkcolor=blue,citecolor=blue]{hyperref}
\usepackage{placeins}
\usepackage{bm} 
\usepackage{nicefrac}
\usepackage{slashed}
\usepackage{marginnote}
\usepackage{combelow}

\newcommand{\ab}[1]{\left\langle#1\right\rangle}

\begin{document}
	
	\preprint{APS/123-QED}

     \title{Quasi-particle hydrodynamics with momentum-dependent relaxation time}

    \author{Arghya Mukherjee}
	\email{arbp.phy@gmail.com}
	\affiliation{Ramakrishna Mission Residential College (Autonomous), Narendrapur, Kolkata-700103, India}
    
    \author{Samapan Bhadury}
	\email{samapan.bhadury@uj.edu.pl}
	\affiliation{Institute of Theoretical Physics, Jagiellonian University ul. St. \L ojasiewicza 11, 30-348 Krakow, Poland}

    \author{Pracheta Singha}
	\email{pracheta.singha@e-uvt.ro}
	\affiliation{Department of Physics, West University of Timişoara, Bulevardul Vasile Pârvan 4, Timişoara 300223, Romania}

    \date{\today}
	
    \begin{abstract}
    We formulate the  relativistic dissipative hydrodynamics of a system of quasi-particles from the  Boltzmann equation within the ambit of relaxation time approximation with modified collision kernels. We focus on two specific scenarios with single quasi-particle species, (i) the extended relaxation time approximation, and (ii) the novel relaxation time approximation. We find that both approaches lead to equivalent results up to first-order in spacetime gradients. We generalize the extended relaxation time approach to incorporate multiple quasi-particle species and obtain the corresponding expressions for the  shear ($\eta_s$) and bulk ($\zeta_s$) viscous coefficients. As an application, we study the temperature dependence of the transport coefficients of hot QCD medium with quasi-gluon and (light and strange) quasi-quark sectors considering the power law ansatz for the momentum dependence of the relaxation time. We explore the impact of the power law exponent on the ratio $\zeta_s/\eta_s$.  Our study suggests that in comparison to a constant exponent, a temperature dependent exponent in the power law ansatz is more suitable for modeling  the  quasi-particle dynamics in the relevant temperature regime of heavy ion collision.         
    \end{abstract}
         
    \date{\today}


\maketitle

\section{Introduction}
\label{intro}
Relativistic viscous hydrodynamics plays a crucial role in  the space-time  evolution of the strongly interacting extreme state of matter produced in the ultra-relativistic heavy ion collisions \cite{Ollitrault:2007du, Florkowski:2010zz, Jaiswal:2016hex, Florkowski:2017olj, Romatschke:2017ejr}. The computational modeling of the collision dynamics from the initial pre-equilibrium stage to the final stage of kinetic freeze-out requires a multi-stage approach of which a major part is the hydrodynamic evolution of the hot and dense quark-gluon plasma (QGP) \cite{Jeon:2015dfa, Florkowski:2017olj}. For the correct description of the dynamical evolution, it is necessary to have a detailed understanding of the transport properties of the  QGP medium. However, precise estimation of transport coefficients based on first principle lattice quantum chromodynamics (LQCD) approach remains challenging \cite{Meyer:2011gj}. On the other hand, in the relevant temperature regime of heavy ion collision experiments, the results from phenomenological modeling \cite{Bernhard:2019bmu, Nijs:2020roc, Parkkila:2021tqq} suggest a significantly smaller specific shear viscosity and a significantly larger specific bulk viscosity compared to the perturbative estimations  \cite{Arnold:2006fz, Arnold:2000dr, Arnold:2003zc}. There has been a conscientious effort to estimate the QGP transport using holographic QCD framework (see Refs.~\cite{Florkowski:2017olj, Schafer:2009dj} for an overview)  as well as effective approaches (see Refs.~\cite{Jeon:1995zm, Sasaki:2008fg, Bluhm:2007nu, Bluhm:2011xu, Chakraborty:2010fr, Plumari:2011mk, Alqahtani:2015qja, Berrehrah:2016vzw, Ozvenchuk:2012kh, Soloveva:2019xph, Marty:2013ita, Lang:2013lla, Czajka:2017bod, Deb:2016myz, Singha:2017jmq, Mykhaylova:2019wci, Mykhaylova:2020pfk} and the references therein) which has enriched our understanding considerably. In the present work we consider the effective kinetic theory approach based on the recent formulation of quasi-particle hydrodynamics from relativistic Boltzmann equation with modified collision kernel \cite{Rocha:2021zcw, Dash:2021ibx}.          

The conventional Anderson and Witting relaxation time approximation (RTA) of the collision kernel \cite{anderson1974relativistic} has been extremely successful in the order by order formulation of the  dissipative hydrodynamics and has been widely used to derive the transport coefficients. However, if the energy (momentum) dependence of the relaxation time is taken into consideration, the microscopic conservation law in the conventional RTA is violated. Similar discrepancy  arises if one considers a general matching condition. Recent formulations based on modified collision kernel \cite{Rocha:2021zcw, Dash:2020zqx, Biswas:2022cla} have addressed this issue. In Ref. \cite{Rocha:2022fqz}, the novel relaxation time approximation (NRTA) has been implemented to study the transport properties of QGP considering a single quasi-particle species with temperature dependent effective mass. The formulation is based on Maxwell-Boltzmann statistics. The quasi-particles in this formulation do not correspond to gluon or quark degrees of freedom. Rather, they correspond to an effective degrees of freedom that satisfies the  QCD thermodynamics. On the other hand, the formulation based on extended relaxation time approximation (ERTA) \cite{Dash:2020zqx} analyze the implications of the power law ansatz in the transport coefficients for both classical and quantum statistics. However, the ERTA formulation does not consider the quasi-particle picture. Thus, to implement either of the frameworks in case of 2+1 flavor QCD, one needs a generalization which we attempt in the present work.

The article is organized as follows: In Sec.~\ref{sec:SSQ} we first discuss the thermodynamic consistency of the single-species quasi-particle model and formulate the corresponding relativistic hydrodynamics from kinetic theory considering two types of collision kernels - (i) ERTA, and (ii) NRTA. Up to first-order in spacetime gradient, we show that both of the collision kernels result in the same out-of-equilibrium corrections to phase-space distribution and hence the transport coefficients. We  obtain expressions for the frame and matching invariant transport coefficients from the entropy production. The results obtained with single quasi-particle species are generalized for the multi-species system in Sec.~\ref{sec:MSQ} with ERTA collision kernel. The multi-species framework is then utilized in Sec.~\ref{sec:R&D} to explore the impact of the power law momentum dependence of the relaxation time on the shear and bulk viscous coefficient of 2+1 flavor QCD medium with quasi-gluon and, light and strange quasi-quark (anti-quark) sectors. 
Finally, in Sec.~\ref{sec:C&O}, we  conclude  with a brief summary and  future outlook.

\textit{Notations and conventions:} Throughout this article, natural units have been adopted i.e. $k_B = \hbar = c = 1$. For metric, the mostly negative convention has been chosen i.e., $g_{\mu\nu} = diag(1, -1, -1, -1)$. A short-hand notation is introduced for the phase-space distribution function, $f (x, p)$, as $f_{\textbf{p}}$ where $p$ is the absolute value of the particle four-momentum, $p^\mu$ i.e. $p^2 \equiv p^\mu p_\mu$ and $x^\mu$ is the spacetime coordinate.. Any macroscopic and microscopic variable at equilibrium will carry a subscript of ‘0’ e.g., $f_{0\textbf{p}}$, whereas the non-equilibrium corrections will be denoted by adding a ‘$\delta$’ in front of the variable e.g., $\delta f_{\textbf{p}}$. Three-vectors will be denoted by bold letters. The scalar product of two Lorentz-vectors is denoted by, $A \cdot B \equiv A^{\mu} B_{\mu}$ where Einstein's summation convention is assumed.

\section{Single Species Quasi-particle}
\label{sec:SSQ}

In this section, we will discuss in detail how to obtain the transport coefficients of a relativistic fluid with single quasi-particle degree of freedom. We will first establish the thermodynamics and then go on to evaluate the transport coefficients from two newly proposed collision kernels, ERTA and NRTA. We will show that when truncated at a first-order, both the collision kernels yield the same result. In the next section, using this, we will determine the transport coefficients of a relativistic fluid with multiple quasi-particle species.

\subsection{Quasi-particle Thermodynamics}
\label{ssec:QT}

To provide an effective kinetic theoretical description of the strongly interacting matter,  we will consider a system with single quasi-particle species without any conserved charge. The effect of the interactions among the quasi-particles is modeled  with a temperature-dependent effective mass, $m(T)$. Such an effective mass allows the incorporation of a desired equation of state (EoS) but destroys the thermodynamic consistency. 
To restore thermodynamic consistency, we covariantly introduce a Bag pressure-like term $B(T)$ into the energy-momentum tensor as \cite{Gorenstein:1995vm, Jeon:1994if, Jeon:1995zm, Chakraborty:2010fr, Romatschke:2011qp, Albright:2015fpa, Alqahtani:2015qja, Tinti:2016bav, Czajka:2017wdo},
\begin{align}
    T^{\mu\nu} = \left(\varepsilon_0 + \delta \varepsilon\right) u^\mu u^\nu - \left(P_0 + \delta P\right) \Delta^{\mu\nu} + 2 h^{(\mu} u^{\nu)} + \pi^{\mu\nu} \equiv \ab{p^\mu p^\nu} + B (T) g^{\mu\nu}, \label{emt-def-KT}
\end{align}
where $\varepsilon_0, P_0$ are the equilibrium energy density and pressure, $\delta \varepsilon, \delta P$ are the out-of-equilibrium corrections to energy density and pressure, $h^\mu$ and $\pi^{\mu\nu}$ are the energy diffusion current and shear viscous pressure, respectively. The fluid four-velocity, $u^\mu$ is normalized to unity as $u\cdot u = 1$, and $\Delta^{\mu\nu} = g^{\mu\nu} - u^\mu u^\nu$ is the projection operator orthogonal to $u^\mu$. The Bag term, ingeneral can have contributions from equilibrium as well as non-equilibrium effects so that we can write, $B(T) = B_0(T) + \delta B(T)$. Here the parenthesis around the indices are defined as, $A^{(\mu} B^{\nu)} = \left(1/2\right) \left(A^\mu B^\nu + A^\nu B^\mu \right)$, whereas the angular brackets represent momentum integral weighted with the non-equilibrium phase-space distribution function $f_{\textbf{p}}$, defined by,
\begin{align}
    \ab{\cdots} \equiv \int_{\textbf{p}} \left(\cdots\right) f_{\textbf{p}}. \label{ang-brak-def}
\end{align} 
We have used a Lorentz invariant integral measure defined as \cite{Romatschke:2011qp, Tinti:2016bav},
\begin{align}
    \int_{\textbf{p}} \equiv \int \frac{g d^3 \textbf{p}}{\left(2\pi\right)^3 E_{\textbf{p}}}
    \label{integral-measure-def}
\end{align}
where $g$ is the degeneracy factor, and $E_{\textbf{p}} = \sqrt{\textbf{p}^2 + m^2}$ is the particle energy in the local rest frame.

We want our system to abide by the thermodynamic relation, $s_0 = \beta \left(\varepsilon_0 + P_0 \right)$, where $s_0$ is the equilibrium entropy density and $\beta = 1/T$ is the inverse temperature. This requires us to impose the condition,
\begin{align}
    \partial_\mu B_0 = - \frac{1}{2} \left( \partial_\mu m^2 \right) \ab{1}_0 \label{thermo-cosist-rel}
\end{align}
where we defined:
\begin{align}
    \ab{\cdots}_0 \equiv \int_{\textbf{p}} \left(\cdots\right) f_{0\textbf{p}},
    \qquad\qquad\qquad
    \ab{\cdots}_1 \equiv \int_{\textbf{p}} \left(\cdots\right) f_{0\textbf{p}} \left(1 - r f_{0\textbf{p}}\right). \label{ang-brak_n-def}
\end{align}
Here $r$ can take values $0, +1, -1$ for particles obeying MB, FD, BE statistics. The equilibrium phase-space distribution function has been defined as,
\begin{align}
    f_{0\textbf{p}} = \left(e^{\beta E_{\textbf{p}}} + r\right)^{-1}. \label{f_0p-def}
\end{align}

Additionally, we can ensure the conservation of energy-momentum tensor by requiring,
\begin{align}
    \partial_\mu B = - \frac{1}{2} \left( \partial_\mu m^2 \right) \ab{1}. \label{emt-conv-rel}
\end{align}
We can coincide the two conditions from Eqs.~\eqref{thermo-cosist-rel} and \eqref{emt-conv-rel} by requiring,
\begin{align}
    \ab{1} = \ab{1}_0, \label{MC}
\end{align}
which can be treated as a matching condition, as we will see in later sections. This matching condition, together with  Eqs.~\eqref{thermo-cosist-rel} and, \eqref{emt-conv-rel} provides $\partial_\mu \delta B (T) = 0$ which allows one to set $\delta B(T) = 0$.

\subsection{\label{ssec:QHRKT}Quasi-particle hydrodynamics and relativistic kinetic theory}

Having fixed the thermodynamics, we focus on building the kinetic theory framework to be used for obtaining the transport properties of the system. We can define the hydrodynamic variables appearing in Eq.~\eqref{emt-def-KT} as moments of the distribution function as,
\begin{subequations}
    \begin{align}
        \varepsilon_0 &= \ab{E_{\textbf{p}}^2}_0 + B_0(T),
        \quad\quad~~
        P_0 = - \ab{\left(1/3\right) \left(p \cdot \Delta \cdot p\right)}_0 - B_0(T), \label{e0,P0-def}\\
        \delta\varepsilon &= \ab{E_{\textbf{p}}^2 \phi_{\textbf{p}}}_1 
        \qquad\qquad\quad~~
        \delta P = - \ab{\left(1/3\right) \left(p \cdot \Delta \cdot p\right) \phi_{\textbf{p}}}_1 
        \label{dele,delP-def}\\
        h^\mu &= \ab{E_{\textbf{p}} p^{\ab{\mu}} \phi_{\textbf{p}}}_1,
        \quad\quad~~~
        \pi^{\mu\nu} = \ab{p^{\langle\mu} p^{\nu\rangle} \phi_{\textbf{p}}}_1, \label{h^m,p^mn-def}
    \end{align}
\end{subequations}
where $A^{\langle\mu} B^{\nu\rangle} = \Delta^{\mu\nu}_{\alpha\beta} A^\alpha B^\beta$, with $\Delta^{\mu\nu}_{\alpha\beta} = \Delta^{(\mu}_\alpha \Delta^{\nu)}_\beta - \left(1/3\right) \Delta^{\mu\nu} \Delta_{\alpha\beta}$ being the rank-4 doubly symmetric, traceless projection operator. The hydrodynamic equations can be obtained from the conservation of energy-momentum tensor i.e. $\partial_\mu T^{\mu\nu} = 0$ by taking projection along and orthogonal to $u_\mu$. These lead to:
\begin{subequations}
    \begin{align}
        D\varepsilon_0 + D \delta \varepsilon + \left(\varepsilon_0 + \delta \varepsilon + P_0 + \delta P \right) \theta - \pi^{\mu\nu} \sigma_{\mu\nu} + \partial_\mu h^\mu + u_\mu D h^\mu &= 0 \label{Heq1}\\
        \left(\varepsilon_0 + \delta \varepsilon + P_0 + \delta P \right) D u^\mu - \nabla^\mu \left(P_0 + \delta P\right) + h^\mu \theta + h^\alpha \Delta^{\mu\nu} \partial_\alpha u_\nu + \Delta^{\mu\nu} D h_\nu + \Delta^{\mu\nu} \partial_\alpha \pi^{\alpha}_\nu &= 0 \label{Heq2}
    \end{align}
\end{subequations}
where the spacetime derivative, `$\partial_\mu$' has been split into two components as $\partial_\mu \equiv u_\mu D + \nabla_\mu$, with $D = u_\mu \partial^\mu$ being the co-moving derivative and $\nabla_\mu = \Delta_\mu^\alpha \partial_\alpha$ being the orthogonal derivative. In the non-relativistic limit/fluid rest frame, the former reduces to time derivative and the latter reduces to spatial derivatives. These two equations also contain the scalar expansion rate, $\theta \equiv \partial_\mu u^\mu$ and fluid stress tensor, $\sigma^{\mu\nu} = \partial^{\langle\mu} u^{\nu\rangle}$.

Using the conservation of energy-momentum tensor from Eq.~\eqref{emt-def-KT} along with Eq.~\eqref{emt-conv-rel} we can arrive at the equation:
\begin{align}
    \int_{\textbf{p}} p^\nu \left[\left(p\cdot \partial\right) f_{\textbf{p}} + \frac{1}{2} \left(\partial_\mu m^2\right) \partial_{(p)}^{\mu} f_{\textbf{p}} \right] = 0. \label{del.T^mn=0}
\end{align}
where $\partial_{(p)}^{\mu} \equiv \partial/\partial p_\mu$ is the derivative with respect to momentum four-vector. With the help of Eq.~\eqref{del.T^mn=0} we can write the relativistic Boltzmann equation as,
\begin{align}
    \left(p\cdot \partial\right) f_{\textbf{p}} + \frac{1}{2} \left(\partial_\mu m^2\right) \partial_{(p)}^{\mu} f_{\textbf{p}} = C[f_{\textbf{p}}], \label{Beq}
\end{align}
where $C[f_{\textbf{p}}]$ is the collision kernel that carries the  information of the interactions among the medium constituents. One can retrieve  the conservation of energy momentum tensor from the first moment of Eq.\eqref{Beq} provided $\int_p p^\mu C[f] = 0$. While the relaxation time approximation (RTA) of the collision kernel  as proposed by Anderson-Witting \cite{anderson1974relativistic} is a widely used approach ~\cite{Tinti:2016bav, Czajka:2017wdo, Strickland:2018ayk, Jaiswal:2019cju, Chattopadhyay:2023hpd} it can not ensure a vanishing first moment of the collision kernel in an arbitrary frame with a momentum-dependent relaxation time. As already mentioned,  successful formulation of relativistic hydrodynamics with momentum-dependent relaxation time has been achieved in recent times using modified collision kernels~\cite{Rocha:2021zcw, Dash:2021ibx}. In the following, we will consider these modified frameworks to obtain the out-of-equilibrium corrections to the phase-space distribution function and the transport coefficients.

\subsection{\label{ssec:ERTA}Extended Relaxation Time Approximation}

The collision kernel of extended relaxation time approximation (ERTA) is given by,
\begin{align}
    C_1[f_{\textbf{p}}] = - \left(E_{\textbf{p}}/\tau_{\rm R}\right) \left(f_{\textbf{p}} - f^*_{0\textbf{p}} \right). \label{CK-ERTA}
\end{align}
where, $\tau_{\rm R} = \tau_{\rm R}(x,p)$ is the relaxation time defining the characteristic timescale in which the particle with four-momentum $p^\mu$ relaxes towards the equilibrium state momentum. A common practice in the existing literature is to parametrize this relaxation time as, $\tau_{\rm R} (x,p) = t_{\rm R}(x) \left(\beta E_{\textbf{p}}\right)^\ell$, where $\ell$ is some exponent determined by the nature of the theory \cite{Dusling:2009df, Dash:2021ibx}.  We also denote $f_{0\textbf{p}}^*$ as the phase-space distribution function at thermodynamic equilibrium and is given by,
\begin{align}
    f^*_{0\textbf{p}} = \left(e^{\beta^*\cdot p} + r\right)^{-1}, \label{f^*_0}
\end{align}
for chargeless massive particles. Here $\beta^*_\mu = u^*_\mu/T^*$ with $u_\mu^*$ and $T^*$ being the fluid four-velocity (normalized to unity as, $u^*\cdot u^* = 1$) and temperature at thermodynamic equilibrium. These thermodynamic variables can be related with the hydrodynamic ones as,
\begin{align}
    u^*_\mu \equiv u_\mu + \delta u_\mu,
    \qquad\qquad
    T^* \equiv T + \delta T. \label{thermo-hydro-rel}
\end{align}
We can expand the thermodynamic equilibrium around the hydrodynamic equilibrium as, $f^*_{0\textbf{p}} = f_{0\textbf{p}} + \delta f_{\textbf{p}}^* = \left(1 + \phi^*_{\textbf{p}} \tilde{f}_{0\textbf{p}} \right) f_{0\textbf{p}}$, which means $\delta f_{\textbf{p}}^* = \phi_{\textbf{p}}^* f_{0\textbf{p}} \tilde{f}_{0\textbf{p}}$, with $\tilde{f}_{0\textbf{p}} \equiv 1 - r f_{0\textbf{p}}$. Here, $r$ controls the statistics of the particle and takes the values, $0, +1, -1$ corresponding to Maxwell-Boltzmann, Fermi-Dirac, and Bose-Einstein statistics respectively.
The non-equilibrium distribution function can also be expanded as, $f_{\textbf{p}} = f_{0\textbf{p}} + \delta f_{\textbf{p}} = \left(1 + \phi_{\textbf{p}} \tilde{f}_{0\textbf{p}} \right) f_{0\textbf{p}}$, which gives $\delta f_{\textbf{p}} = \phi_{\textbf{p}} f_{0\textbf{p}} \tilde{f}_{0\textbf{p}}$. Using
\begin{align}
    \left(p\cdot \partial\right) f_{0\textbf{p}} + \frac{1}{2} \partial_\mu m^2 \left(\partial_{(p)}^\mu f_{0\textbf{p}}\right) = \left(A_{\textbf{p}} \theta - \beta p^\mu p^\nu \sigma_{\mu\nu} \right) f_{0\textbf{p}} \tilde{f}_{0\textbf{p}}, \label{ERTA_Beq-LHS}
\end{align}
we can re-express the Boltzmann equation as,
\begin{align}
    \left(A_{\textbf{p}} \theta - \beta p^\mu p^\nu \sigma_{\mu\nu} \right) f_{0\textbf{p}} \tilde{f}_{0\textbf{p}} = - \left(E_{\textbf{p}}/\tau_{\rm R}\right) \left(\phi_{\textbf{p}} - \phi_{\textbf{p}}^*\right) f_{0\textbf{p}} \tilde{f}_{0\textbf{p}}, \label{Beq-ERTA}
\end{align}
where,
\begin{align}
    A_{\textbf{p}} = - \beta \left[ \left(c_s^2 - \frac{1}{3}\right) E_{\textbf{p}}^2 + \frac{m^2}{3} + \frac{\beta c_s^2}{2} \frac{\partial m^2}{\partial\beta}\right]. \label{A_p-def}
\end{align}
 To obtain Eqs.~\eqref{ERTA_Beq-LHS} and \eqref{A_p-def}, we have replaced the comoving derivatives in terms of other gradients  using the relations
\begin{align}
    \dot{\beta} = \beta c_s^2 \theta,
    \qquad\quad
    \dot{u}^\mu = - \nabla^\mu \ln{\beta}, \label{Dbeta-&-Du}
\end{align}
that follow from the hydrodynamic  equations \eqref{Heq1} and \eqref{Heq2}. Here, $c_s^2$ in Eq.~\eqref{Dbeta-&-Du} is the square of the speed of sound, given by,
\begin{align}
    c_s^2 = \frac{J_{31}}{J_{30} + \frac{\beta}{2} \frac{\partial m^2}{\partial\beta} J_{10}}. \label{c_s^2}
\end{align}
In order to derive these relations, we have used two types of thermodynamic integrals defined as
\begin{subequations}
    \begin{align}
        I_{nq} &\equiv \frac{\left(-1\right)^q}{\left(2q+1\right)!!} \left\langle E_{\textbf{p}}^{n-2q} \left( p \cdot \Delta \cdot p\right)^q \right\rangle_0, \label{I_nq}\\
        J_{nq} &\equiv \frac{\left(-1\right)^q}{\left(2q+1\right)!!} \left\langle E_{\textbf{p}}^{n-2q} \left(p \cdot \Delta \cdot p \right)^q \right\rangle_1 \label{J_nq}.
    \end{align}
\end{subequations}
For further details on the properties of these momentum integrals see Appendix-\ref{TIP}. Keeping terms up to first-order in spacetime gradient we write,
\begin{align}
    \phi^*_{\textbf{p}} = \beta \Big[ - p_\mu \delta u^\mu + \beta E_{\textbf{p}} \delta T\Big].
    \label{phi^*_1}
\end{align}
To determine $\phi_{\textbf{p}}$, we expand it using a set of orthogonal polynomial, $P_n^{(l)}$ as \cite{Denicol:2012cn},
\begin{align}
    \phi_{\textbf{p}} = \sum_{n,l = 0}^\infty \Phi_n^{\ab{\mu_1\cdots\mu_l}} p_{\langle\mu_1} \cdots p_{\mu_l\rangle} P_n^{(l)}, \label{phi-expansion_erta}
\end{align}
where $p^{\langle\mu_1} \cdots p^{\mu_l\rangle} = \Delta^{\mu_1\cdots\mu_l}_{\alpha_1\cdots\alpha_l} p^{\alpha_1} \cdots p^{\alpha_l}$ are the irreducible tensors for $l = 0,1,\cdots$, with $\Delta^{\mu_1\cdots\mu_l}_{\alpha_1\cdots\alpha_l}$ defined in Ref.~\cite{DeGroot:1980dk}. The label `$l$' corresponds to the rank of the term under consideration, e.g. $l = 0, 1, 2, \cdots$ correspond to scalars, vectors, rank-2 tensors, and so on. The label `$n$' corresponds to the $n$-th term of some rank-$q$ quantity. The polynomials $P_n^{(l)}$ can be determined from the Gram-Schmidt orthogonalization process. In the present case, we have chosen $P_0^{(0)} = \beta E_{\textbf{p}}$ and $P_0^{(l\geq1)} = 1$. Apparently it seems straightforward to obtain the coefficients, $\Phi_n^{\ab{\mu_1\cdots\mu_l}}$ by solving the Boltzmann equation given in Eq.\eqref{Beq-ERTA}. 
However, it should be noted that the
complete determination of $\Phi_n^{\ab{\mu_1\cdots\mu_l}}$ will require the knowledge of $\phi_{\textbf{p}}^*$  given in Eq.\eqref{phi^*_1}. At this point we need to introduce the matching and frame conditions. The matching and frame conditions respectively fix the definitions of temperature and  fluid four-velocity   which essentially  determines $\delta u^\mu$ and $\delta T$ given in Eq.\eqref{phi^*_1} and hence completely specify $\phi_{\textbf{p}}^*$. In this work we will use the following matching and frame conditions,
\begin{align}
    \ab{g_{\textbf{p}} \phi_{\textbf{p}}}_1 = 0,
    \qquad\quad\quad
    \ab{q_{\textbf{p}} p^{\ab{\mu}} \phi_{\textbf{p}}}_1= 0, \label{mcfc-erta}
\end{align}
which gives us\footnote{In Eq.~\eqref{a,b_p-erta_mu0} we have used general matching condition but for the present work, we will set $g_{\textbf{p}}=1$.},
\begin{align}
    \delta T = \frac{\ab{g_{\textbf{p}} \tau_{\rm R} A_{\textbf{p}}/E_{\textbf{p}}}_1 \theta}{\beta^2 \ab{g_{\textbf{p}} E_{\textbf{p}}}_1},
    \qquad\qquad\qquad
    \delta u^\mu = 0. \label{a,b_p-erta_mu0}
\end{align}
We can use these in the Boltzmann equation, \eqref{Beq-ERTA} to determine $\Phi_n^{\ab{\mu_1\cdots\mu_l}}$ of $l$-th rank, by multiplying both sides of Eq.~\eqref{Beq-ERTA} by $P_q^{(l)} p^{\langle\mu_1} \cdots p^{\mu_l\rangle}$ and integrating over momentum with the measure defined in Eq.~\eqref{integral-measure-def}. This allows us to write,\footnote{Note that we obtained $\Phi_n^{\ab{\alpha}} = 0$ in Eq.~\eqref{Phi_n^mu-erta} as the system under consideration consists of quasi-particles of zero chemical potential, giving us $\delta u^\mu = 0$. Furthermore, it is possible to show $\Phi_0 = \beta\, \delta T$ and $\Phi_0^{\ab{\mu}} = - \beta \delta u^\mu$. This fact indicates that $\phi^*_{\textbf{p}}$ is  related to the homogeneous part of the solution to Boltzmann equation.}

\begin{subequations}
    \begin{alignat}{2}
        \Phi_n &= \frac{1}{A_{n}^{(0)}} \left[ \beta^2 \delta T \ab{P_n^{(0)} E_{\textbf{p}}^2/\tau_{\rm R}}_1 - \ab{A_{\textbf{p}} P_n^{(0)}}_1 \theta \right],
        \qquad\qquad\qquad &{\rm for~n\geq0}\label{Phi_n-erta}\\
        \Phi_n^{\ab{\mu_1}} &= 0, 
        \qquad\qquad\qquad &{\rm for~n\geq0} \label{Phi_n^mu-erta}\\
        \Phi_n^{\ab{\mu_1\mu_2}} &= \frac{2\beta}{15 A_n^{(2)}} \ab{\left(p\cdot\Delta\cdot p\right)^2 P_n^{(2)}}_1 \sigma^{\mu_1\mu_2}, 
        \qquad\qquad\qquad &{\rm for~n\geq0} \label{Phi_m^munu-erta}\\
        \Phi_n^{\ab{\mu_1\cdots\mu_l}} &= 0, 
        \qquad\qquad\qquad &{}{\rm for}{}~n\geq0, \ell\geq3 \label{Phi_n^mu1-l-erta}
    \end{alignat}
\end{subequations}
To obtain the above result, we have used the orthogonality property of $P_n^{(l)}$,
\begin{align}
    \frac{l!}{\left(2 l + 1\right)!!} \ab{\left(E_{\textbf{p}}/\tau_{\rm R}\right) \left(p \cdot \Delta \cdot p\right)^l P_n^{(l)} P_q^{(l)}}_1 = A^{(l)}_{n} \delta_{nq}. \label{P_n-orthogonality}
\end{align}
where $A_n^{(l)}$ is a normalization constant and we will treat Eq.~\eqref{P_n-orthogonality} as its definition after setting $n=q$. The orthogonal polynomials allow us to express any function of phase-space variables, $\mathcal{G}_{\textbf{p}}$ as,
\begin{align}
    \mathcal{G}_{\textbf{p}} = \sum_{n=0}^\infty \frac{1}{A_n^{(l)}} \frac{l!}{\left(2 l + 1\right)!!} \ab{\left(E_{\textbf{p}}/\tau_{\rm R}\right) \mathcal{G}_{\textbf{p}} P_{n}^{(l)} \left(p\cdot\Delta\cdot p\right)^l}_1 P_n^{(l)}. \label{G_p}
\end{align}
Using this property we can substitute Eqs.~\eqref{Phi_n-erta}-\eqref{Phi_n^mu1-l-erta} into Eq.~\eqref{phi-expansion_erta} to obtain,
\begin{align}
    \phi_{\textbf{p}} = \beta^2 E_{\textbf{p}} \delta T - \left(\frac{\tau_{\rm R} A_{\textbf{p}}}{E_{\textbf{p}}}\right) \theta + \left(\frac{\beta \tau_{\rm R}}{E_{\textbf{p}}}\right) p_{\langle\alpha} p_{\beta\rangle}\, \sigma^{\alpha\beta}. \label{phi_p-erta}
\end{align}
With this, we can determine the out-of-equilibrium correction to distribution, $\delta f_{\textbf{p}}$ for ERTA and hence the transport coefficients. Our next task is to determine the same for the case of novel relaxation time approximation (NRTA).

\subsection{\label{ssec:NRTA}Novel Relaxation Time Approximation}

In this section, we use a collision kernel inspired from Ref.~\cite{Rocha:2022fqz}, which conserves the energy-momentum tensor by construction. We have re-derived the collision kernel of NRTA to be suitable for quantum statistics. The linear collision kernel\footnote{Note that, the linearized collision kernel can be related to a linear collision operator, $\hat{L}$ such that, $C_2[f_{\textbf{p}}] = f_{0\textbf{p}} \tilde{f}_{0\textbf{p}} \hat{L} \phi_{\textbf{p}}$. In the case of uncharged particles, $\hat{L}$ can have 4 degenerate eigenmodes with zero eigenvalues. Therefore, we may approximate this as,
\begin{align*}
    \hat{L} \propto \mathbb{1} - \sum_{n=0}^{3} \ket{\lambda_n}\bra{\lambda_n}.
\end{align*}
}
is found to be:
\begin{align}
    C_2[f_{\textbf{p}}] = - \frac{E_{\textbf{p}}}{\tau_\mathrm{R}} \left[\phi_{\textbf{p}} - \frac{\ab{\left(E_{\textbf{p}}^2/\tau_{\rm R}\right) \phi_{\textbf{p}}}_1 E_{\textbf{p}}}{\ab{\left(E_{\textbf{p}}^3/\tau_{\rm R}\right)}_1} - \frac{\ab{\left(E_{\textbf{p}}/\tau_{\rm R}\right) \phi_{\textbf{p}}\, p^{\ab{\mu}}}_1 p_{\ab{\mu}}}{\ab{\left(1/3\right) \left(E_{\textbf{p}}/\tau_{\rm R}\right) \left(p \cdot \Delta\cdot p\right)}_1} \right] f_{0\textbf{p}} \tilde{f}_{0\textbf{p}} \label{QSNRTA}
\end{align}
Similar to section \ref{ssec:ERTA}, to solve for $\phi_{\textbf{p}}$, we will expand $\phi_{\textbf{p}}$ according to Eq.~\eqref{phi-expansion_erta}. Substituting Eq.~\eqref{QSNRTA} in Eq.~\eqref{Beq} we can write the Boltzmann equation under NRTA with the help of Eq.~\eqref{ERTA_Beq-LHS} as,
\begin{align}
    \left( A_{\textbf{p}} \theta - \beta p^\mu p^\nu \sigma_{\mu\nu} \right) f_{0\textbf{p}} \tilde{f}_{0\textbf{p}} = - \frac{E_{\textbf{p}}}{\tau_\mathrm{R}} \left[\phi_{\textbf{p}} - \frac{\ab{\left(E_{\textbf{p}}^2/\tau_{\rm R}\right) \phi_{\textbf{p}}}_1 E_{\textbf{p}}}{\ab{\left(E_{\textbf{p}}^3/\tau_{\rm R}\right)}_1} - \frac{\ab{\left(E_{\textbf{p}}/\tau_{\rm R}\right) \phi_{\textbf{p}}\, p^{\ab{\mu}}}_1 p_{\ab{\mu}}}{\ab{\left(1/3\right) \left(E_{\textbf{p}}/\tau_{\rm R}\right) \left(p \cdot \Delta\cdot p\right)}_1} \right] f_{0\textbf{p}} \tilde{f}_{0\textbf{p}} \label{Beq-NRTA}
\end{align}
where $A_{\textbf{p}}$ is defined in Eq.~\eqref{A_p-def}. Upon substitution of Eq.~\eqref{phi-expansion_erta}, one finds $\Phi_0$ and $\Phi_0^\mu$ gets removed from the equation by virtue of the second and third terms in Eq.~\eqref{Beq-NRTA}. 
Therefore, they cannot be determined from the Boltzmann equation, rather we have to use the matching and frame conditions. Thus, using Eq.~\eqref{phi-expansion_erta} in Eq.~\eqref{mcfc-erta} and setting $g_{\textbf{p}} = 1$ we find,
\begin{subequations}
    \begin{align}
        \Phi_0 &= - \sum_{n = 1}^\infty \Phi_n \frac{\left\langle P_n^{(0)}\right\rangle_1}{\left\langle P_0^{(0)}\right\rangle_1}, \label{Phi_0-nrta}\\
        \Phi_0^{\ab{\mu}} &= 0. \label{Phi_0^mu-nrta}
\end{align}
\end{subequations}

Thus, Eqs.~\eqref{Phi_0-nrta} and \eqref{Phi_0^mu-nrta} describe the $n=0$ elements of $\Phi_n^{\langle\mu_1\cdots\mu_\ell\rangle}$ for $\ell = 0,1$ i.e. the homogeneous part of the solution of Eq.~\eqref{Beq-NRTA} (similar to $\phi_{\textbf{p}}^*$ in the case of ERTA). To determine rest of the $\Phi_n^{\langle\mu_1\cdots\mu_\ell\rangle}$, we substitute Eq.~\eqref{phi-expansion_erta} into Eq.~\eqref{Beq-NRTA} to obtain,
\begin{subequations}
    \begin{alignat}{2}
        \Phi_n  &= - \frac{\left\langle A_{\textbf{p}} P_n^{(0)} \right\rangle_1 \!\! \theta}{A_n^{(0)}},
        \quad~~\qquad\qquad\qquad\qquad\qquad~~\,\quad
        &{}{\rm for}&{}~n\geq1 \label{Phi-scalar-nrta}\\
        \Phi^{\langle\mu\rangle}_n &= 0,
        &{}{\rm for}&{}~n\geq1 \label{Phi-vector-nrta}\\
        \Phi_n^{\langle\mu\nu\rangle} &= \frac{2 \beta}{15\, A_{n}^{(2)}} \left\langle \left(p \cdot \Delta \cdot p\right)^2 P_n^{(2)} \right\rangle_1 \sigma^{\mu\nu},
        &{}{\rm for}&{}~n\geq0 \label{Phi-tensor-nrta}\\
        \Phi_n^{\langle\mu_1\cdots\mu_l\rangle} &= 0.
        &{}{\rm for}&{}~n\geq0, l\geq3. \label{Phi-rank>2-nrta}
    \end{alignat}
\end{subequations}
This fixes all the unknown coefficients $\Phi_n^{\ab{\mu_1\cdots \mu_l}}$ and hence, we are ready to write down the expression of $\phi_{\textbf{p}}$. Examining Eqs.~\eqref{Phi_0-nrta}-\eqref{Phi-rank>2-nrta} we can conclude that $\phi_{\textbf{p}}$ can have the following structure,
\begin{align}
    \phi_{\textbf{p}} = F_{\textbf{p}}^{(0)} \theta + F_{\textbf{p}}^{(2)} p^{\langle\mu} p^{\nu\rangle} \sigma_{\mu\nu}, \label{phi_1p-structure}
\end{align}
where $F_{\textbf{p}}^{(0)}$ and $F_{\textbf{p}}^{(2)}$ are the coefficients of scalar and tensor structures. Recalling Eqs.~\eqref{phi-expansion_erta} and, \eqref{Phi_0-nrta}-\eqref{Phi-rank>2-nrta} we can write,
\begin{subequations}
    \begin{align}
        F_{\textbf{p}}^{(0)} = \sum_{n = 0}^\infty \frac{\left\langle A_{\textbf{p}} P_n^{(0)} \right\rangle_1}{A_n^{(0)}} \left( \frac{\left\langle P_n^{(0)}\right\rangle_1}{\left\langle P_0^{(0)}\right\rangle_1} P_0^{(0)} - P_n^{(0)} \right), \label{F_p^0}\\
        F_{\textbf{p}}^{(2)} = \sum_{n=0}^\infty \frac{2 \beta}{15\, A_{n}^{(2)}} \left\langle \left(p \cdot \Delta \cdot p\right)^2 P_n^{(2)} \right\rangle_1 P_n^{(2)}. \label{F_p^2}
    \end{align}
\end{subequations}
where we note the $n=0$ term on the right-hand side of Eq.~\eqref{F_p^0} is zero.
To simplify these expressions, as in the case of ERTA, we will use Eq.~\eqref{G_p}. By choosing $\mathcal{G}_{\textbf{p}} = A_{\textbf{p}} \tau_{\rm R}/\left(u\cdot p\right)$ for $l = 0$ we can write, $A_{\textbf{p}} \tau_{\rm R}/\left(u\cdot p\right) = \sum_{n=0}^\infty \frac{1}{A_n^{(0)}} \left\langle A_{\textbf{p}} P_n^{(0)} \right\rangle_1 P_{n}^{(0)}$ and consequently, $\ab{A_{\textbf{p}} \tau_{\rm R}/E_{\textbf{p}}}_1 = \sum_{n=0}^\infty \left\langle A_{\textbf{p}} P_n^{(0)} \right\rangle_1 \ab{P_{n}^{(0)}}_1/A_n^{(0)}$. Therefore we can write $F_{\textbf{p}}^{(0)}$ as,
\begin{align}
    F_{\textbf{p}}^{(0)} &= \ab{\left(\frac{A_{\textbf{p}} \tau_{\rm R}}{E_{\textbf{p}}}\right)}_1 \frac{E_{\textbf{p}}}{J_{10}} - \left(\frac{A_{\textbf{p}} \tau_{\rm R}}{E_{\textbf{p}}}\right). \label{F_p^0-nrta}
\end{align}
And, similarly, choosing $\mathcal{G}_{\textbf{p}} = \tau_{\rm R}/\left(u\cdot p\right)$ with $l = 2$, we can write $F_{\textbf{p}}^{(0)}$ as,
\begin{align}
    F_{\textbf{p}}^{(2)} &= \left(\frac{\beta \tau_{\rm R}}{E_{\textbf{p}}}\right). \label{F_p^2-nrta}
\end{align}
Thus, we can use Eqs.~\eqref{F_p^0-nrta} and \eqref{F_p^2-nrta} into Eq.~\eqref{phi_1p-structure} to obtain the solution of the Boltzmann equation with the collision kernel of Eq.~\eqref{QSNRTA} as,
\begin{align}
    \phi_{\textbf{p}} = \left\{\ab{\frac{A_{\textbf{p}} \tau_{\rm R}}{\left(u\cdot p\right)}}_1 \frac{\left(u\cdot p\right)}{J_{10}} - \frac{A_{\textbf{p}} \tau_{\rm R}}{\left(u\cdot p\right)} \right\}\theta + \frac{\beta \tau_{\rm R}}{\left(u\cdot p\right)} p^{\langle\mu} p^{\nu\rangle} \sigma_{\mu\nu}. \label{phi_p-nrta}
\end{align}
Interestingly, we note that the phase-space correction, $\phi_{\textbf{p}}$ for the case of ERTA with the matching condition $g_{\textbf{p}} = 1$ in \eqref{mcfc-erta} gives us exactly Eq.~\eqref{phi_p-nrta} by using $\delta T$ from Eq.~\eqref{a,b_p-erta_mu0}. In the next part, we will use this information to obtain the transport coefficients.

\subsection{\label{ssec:TC}Transport Coefficients}
%
In the first-order theory of uncharged fluids, the dissipative currents are given by Navier-Stokes equations as
\begin{align}
    \delta \varepsilon = e \theta,
    \qquad\qquad
    \delta P = \rho \theta,
    \qquad\qquad
    h^\mu = \kappa \nabla^\mu \alpha,
    \qquad\qquad
    \pi^{\mu\nu} = 2 \eta \sigma^{\mu\nu}, \label{NS-Eqs}
\end{align}
where $e, \rho, \kappa$ and $\eta$ are the transport coefficients, although only two frame independent transport coefficients exist that gives rise to dissipation. To determine the transport coefficients appearing in Eq.~\eqref{NS-Eqs}, we substitute Eq.~\eqref{phi_1p-structure} using \eqref{F_p^0-nrta} and \eqref{F_p^2-nrta} into Eqs.~\eqref{dele,delP-def} and \eqref{h^m,p^mn-def} and compare with Eq.~\eqref{NS-Eqs} to find,
\begin{subequations}
    \begin{align}
        e &= \ab{\left(A_{\textbf{p}} \tau_{\rm R}/E_{\textbf{p}}\right)}_1 \left(J_{30}/J_{10}\right) - \ab{A_{\textbf{p}} \tau_{\rm R} E_{\textbf{p}}}_1, \label{e}\\
        \rho &= \ab{\left(A_{\textbf{p}} \tau_{\rm R}/E_{\textbf{p}}\right)}_1 \left(J_{31}/J_{10}\right) + \ab{\left(1/3\right) \left(A_{\textbf{p}} \tau_{\rm R}/E_{\textbf{p}}\right) \left(p\cdot\Delta\cdot p\right)}_1, \label{rho}\\
        \kappa &= 0 \label{kappa}\\
        \eta &= \frac{\beta}{15} \ab{\left(p\cdot\Delta\cdot p\right)^2 \left(\tau_{\rm R}/E_{\textbf{p}}\right)}_1. \label{eta}
    \end{align}
\end{subequations}
The fact that here we have, $\kappa = 0$ can be understood from the absence of any vector gradient in the expressions of $\phi_{\textbf{p}}$ i.e. Eqs.~\eqref{phi_p-erta} and \eqref{phi_p-nrta}. To determine the matching and frame invariant transport coefficients, we employ the second law of thermodynamics i.e. we will determine the divergence of entropy four-current. From Boltzmann's H-theorem we can write the entropy four-current of particles obeying quantum statistics to be given by,
\begin{align}
    S^\mu = - \int_{\textbf{p}} p^\mu \left(f_{\textbf{p}} \ln{f_{\textbf{p}}} + r \tilde{f}_{\textbf{p}} \ln{\tilde{f}_{\textbf{p}}}\right) \label{QSNRTA-S^mu}
\end{align}
Taking the four-divergence of Eq.~\eqref{QSNRTA-S^mu} 
and keeping terms up to quadratic order in spacetime gradient we can write,
\begin{align}
    \partial_\mu S^\mu = - \int_{\textbf{p}} C[f_{\textbf{p}}] \phi_{\textbf{p}}. \label{d.S1}
\end{align}
Now we substitute the expression of $\phi_{\textbf{p}}$ from Eq.~\eqref{phi-expansion_erta} and for the collision kernel, we use both of ERTA and NRTA from Eqs.~\eqref{CK-ERTA} and \eqref{QSNRTA}. Using appropriate expressions of $\Phi_n^{\ab{\mu_1\cdots\mu_\ell}}$ for both the cases, we arrive at,
\begin{align}
    \partial_\mu S^\mu = \sum_{n=1}^\infty \Phi_n^2 A_n^{(0)} + \sum_{n=1}^\infty \Phi_n^{\ab{\mu}} \Phi_{n,\,\ab{\mu}} A_n^{(1)} + \sum_{n=0}^\infty \Phi_n^{\ab{\mu_1 \mu_2}} \Phi_{n,\,\ab{\mu_1 \mu_2}} A_n^{(2)} + \sum_{l=3}^\infty \sum_{n=0}^\infty \Phi_n^{\ab{\mu_1 \cdots\mu_l}} \Phi_{n,\,\ab{\mu_1 \cdots \mu_{l}}} A_n^{(l)} \label{del.S3}
\end{align}
Following the same argument as in Ref.~\cite{Rocha:2022fqz}, we may conclude that the right-hand side of Eq.~\eqref{del.S3} is positive. Then, recalling the relation \eqref{G_p} and using that in Eq.~\eqref{del.S3} along with Eqs.~\eqref{Phi_n-erta}-\eqref{Phi_n^mu1-l-erta} and \eqref{Phi-scalar-nrta}-\eqref{Phi-rank>2-nrta} we can write,
\begin{align}
    \partial_\mu S^\mu = \beta \zeta_s \theta^2 + 2 \beta \eta_s \sigma^{\mu\nu} \sigma_{\mu\nu} \label{del.S-final}
\end{align}
where,
\begin{subequations}
    \begin{align}
        \zeta_s &= \frac{1}{\beta} \ab{\left(\tau_{\rm R}/E_{\textbf{p}}\right) A_{\textbf{p}}^2}_1, \label{zeta_s}\\
        \eta_s &= \frac{\beta}{15} \ab{\left(\tau_{\rm R}/E_{\textbf{p}}\right) \left(p\cdot\Delta\cdot p\right)^2}_1, \label{eta_s}
    \end{align}
\end{subequations}
are the transport coefficients producing entropy and are frame invariant.

\section{Multi-species Quasiparticle}
\label{sec:MSQ}

In the previous section, we have established that up to first-order in spacetime gradients, the two modified collision kernels of ERTA and NRTA, produce identical results. Therefore, in the present section, we will only work with the ERTA framework due to its relatively simpler structure and discuss the scenario with  multiple quasi-particle species.

\subsection{Quasiparticle Thermodynamics}
\label{ssec:QT_j}

We  consider a system with `$N$' number of quasi-particle species. We  denote the phase-space distribution function of the $j$-th species by $f^{(j)}_{\textbf{p}} = f^{(j)}_{0\textbf{p}} + \delta f^{(j)}_{\textbf{p}}$, where $f^{(j)}_{0\textbf{p}}$ corresponds to the equilibrium part and $\delta f^{(j)}_{\textbf{p}} = \phi^{(j)}_{\textbf{p}} f^{(j)}_{0\textbf{p}} \tilde{f}^{(j)}_{0\textbf{p}}$ represents the out-of-equilibrium correction where $\tilde{f}^{(j)}_{0\textbf{p}} = 1 - r_j f^{(j)}_{0\textbf{p}}$. The parameter $r_j$  controls the statistics of the $j$-th particle species and takes the discrete values, $0, +1, -1$ corresponding to Maxwell-Boltzmann, Fermi-Dirac, and Bose-Einstein statistics respectively. Then we may define the brackets for multi-species as,
\begin{align}
    \ab{\left(\cdots\right)}^{(j)} \equiv \int_{\textbf{p}_j} \left(\cdots\right) f_{\textbf{p}}^{(j)},
    \qquad
    \ab{\left(\cdots\right)}_{0}^{(j)} \equiv \int_{\textbf{p}_j} \left(\cdots\right) f_{0\textbf{p}}^{(j)},
    \qquad
    \ab{\left(\cdots\right)}_{1}^{(j)} \equiv \int_{\textbf{p}_j} \left(\cdots\right) f_{0\textbf{p}}^{(j)} \left(1 - r f_{0\textbf{p}}^{(j)}\right)
    \label{<>^j-def}
\end{align}
where the integral is defined as,
\begin{align}
    \int_{\textbf{p}_j} \equiv \int \frac{g_j d^3 \textbf{p}_j}{\left(2\pi\right)^3 E_{\textbf{p}_j}}, \label{int-meas_j-def}
\end{align}
where $g_j$ is the degeneracy factor of $j$-th species of particle, and $E_{\textbf{p}_j} = \sqrt{\textbf{p}_j^2 + m_j^2}$ is the particle energy of $j$-th species in the local rest frame with $m_j$ being the temperature-dependent mass of the $j$-th particle. From this point onward, we will drop the subscript, `$j$' in momenta (except in Eq.~\eqref{phi^*_1:j}) and its species dependence can be understood from the context. While the tensor decomposition of the energy-momentum tensor remains the same as Eq.~\eqref{emt-def-KT}, the kinetic theory definition is modified to,
\begin{align}
    T^{\mu\nu} = \sum_{j=1}^N \ab{p^\mu p^\nu}^{(j)} + B(T) g^{\mu\nu}, \label{T^mn-def_j}
\end{align}
where thermodynamic consistency requires the Bag pressure to satisfy the relation,
\begin{align}
    \left(\partial^\nu B_0\right) = - \frac{1}{2} \sum_j \left(\partial^\nu m_{j}^2\right) \ab{1}^{(j)}_{0}. \label{sum-Bag-eom}
\end{align}
In this case of multi-species quasiparticles, we demand the following matching condition,
\begin{align}
    \sum_j \left(\partial^\nu m_{(j)}^2\right) \ab{1}^{(j)} = \sum_j \left(\partial^\nu m_{(j)}^2\right) \ab{1}^{(j)}_{0}. \label{sum-matching-condition}
\end{align}
Similar to the single-species quasiparticle case, such a matching condition can be used to show that the out-of-equilibrium part of the Bag pressure vanishes.

\subsection{Quasiparticle hydrodynamics and relativistic kinetic theory}
\label{ssec:QH-RKT_j}

Consequently, the components of $T^{\mu\nu}$ are given by,
\begin{subequations}
    \begin{align}
        \varepsilon_0 &= \sum_{j=1}^{N} \ab{E_{\textbf{p}}^2}_0^{(j)} + B_0(T),
        \qquad\quad
        P_0 = - \sum_{j=1}^{N} \ab{\left(1/3\right) \left(p \cdot \Delta \cdot p\right)}_0^{(j)} - B_0(T), \label{e0,P0-def_j}\\
        \delta\varepsilon &= \sum_{j=1}^{N} \ab{E_{\textbf{p}}^2 \phi_{\textbf{p}}^{(j)}}_1^{(j)},
        \qquad\qquad\quad
        \delta P = - \sum_{j=1}^{N} \ab{\left(1/3\right) \left(p \cdot \Delta \cdot p\right) \phi_{\textbf{p}}^{(j)}}_1^{(j)}, \label{dele,delP-def_j}\\
        h^\mu &= \sum_{j=1}^{N} \ab{E_{\textbf{p}} p^{\ab{\mu}} \phi_{\textbf{p}}^{(j)}}_1^{(j)},
        \qquad\quad~
        \pi^{\mu\nu} = \sum_{j=1}^{N} \ab{p^{\langle\mu} p^{\nu\rangle} \phi_{\textbf{p}}^{(j)}}_1^{(j)}, \label{h^m,p^mn-def_j}
    \end{align}
\end{subequations}
where the correction to the phase-space distribution function of the $j$-th species i.e. $\phi_{\textbf{p}}^{(j)}$ can be determined from the Boltzmann's equation for $j$-th species given by,
\begin{align}
    p^\mu \partial_\mu f^{(j)}_{\textbf{p}} + \frac{1}{2} \left(\partial_\mu m^2_{j}\right) \left(\partial^\mu_{(p)} f^{(j)}_{\textbf{p}}\right) = - \left(E_{\textbf{p}}/\tau_{\rm R}^{(j)}\right) \left(f^{(j)}_{\textbf{p}} - f^{(j)*}_{0\textbf{p}} \right). \label{Beq-ERTA_j}
\end{align}
where we have used the extended relaxation time approximation for the collision kernel \footnote{In principle, the relaxation time of the $j$-th particle species, $\tau_{\rm R}^{(j)}$ should be different for each species. However, following the arguments in Ref.~\cite{Bhadury:2020ngq} it may be shown that we end up having $\tau_{\rm R}^{(j)} = \tau_{\rm R}$ for all $j$ species and this is used from this point onward. The study of the more general case where the relaxation times of each species can be different is an interesting topic and is left for future investigations.}. Here we have defined,
\begin{align}
    f_{0\textbf{p}}^{(j)*} = \left(e^{\beta^*\cdot p} + r_j\right)^{-1}, \label{f_0p^*-j}
\end{align}
as the equilibrium distribution function in terms of thermodynamic variables (starred ones). The unstarred quantities correspond to hydrodynamic variables.  As in Refs.~\cite{Dash:2020zqx, Dash:2021ibx} we can Taylor expand each of the `$N$'-types of distribution functions (starred) as, $f_{0\textbf{p}}^{(j)*} = f_{0\textbf{p}}^{(j)} + \delta f_{\textbf{p}}^{(j)*}$. Expressing the correction part as, $\delta f_{\textbf{p}}^{(j)*} = \phi_{\textbf{p}}^{(j)*} f_{0\textbf{p}}^{(j)}$ the Taylor expansion mentioned in the previous line allows us to write,
\begin{align}
    \phi^{(j)*}_{\textbf{p}} = \beta \Big[ - p_{j}^\mu \delta u_\mu + \beta E_{\textbf{p}_j} \delta T\Big].
    \label{phi^*_1:j}
\end{align}
Recalling the relation, $\delta f_{\textbf{p}}^{(j)} = \phi_{\textbf{p}}^{(j)} f_{0\textbf{p}}^{(j)}$ we can re-write the Boltzmann's equation up to first-order in spacetime gradient as,
\begin{align}
    \left(A^{(j)}_{\textbf{p}}\, \theta - \beta p^\mu p^\nu \sigma_{\mu\nu} \right) f^{(j)}_{0\textbf{p}} \tilde{f}^{(j)}_{0\textbf{p}} = - \left(E_{\textbf{p}}/\tau_{\rm R} \right) \left(\phi_{\textbf{p}}^{(j)} - \phi_{\textbf{p}}^{(j)*}\right) f_{0\textbf{p}}^{(j)} \tilde{f}_{0\textbf{p}}^{(j)}, \label{Beq-ERTA_j2}
\end{align}
where, 
\begin{align}
    A_{\textbf{p}}^{(j)} = - \beta \left\{E_{\textbf{p}}^2 \left(c_s^2 - \frac{1}{3}\right) + \frac{m_j^2}{3} + \frac{\beta c_s^2}{2} \left(\frac{\partial m_{j}^2}{\partial\beta}\right)\right\}. \label{A_p^j-def}
\end{align}
The expression for the speed of sound square of the `$N$' species fluid is modified to,
\begin{align}
    c_s^2 = \frac{\sum_{j=1}^N J^{(j)}_{31}}{\sum_{j=1}^N \left[\frac{\beta}{2} \left(\frac{\partial m_{j}^2}{\partial \beta}\right) J^{(j)}_{10} + J^{(j)}_{30}\right]}, \label{sum-c_s^2:j}
\end{align}
where the thermodynamic integral for the $j$-th species is defined as,
\begin{subequations}
    \begin{align}
        I_{nq}^{(j)} &\equiv \frac{\left(-1\right)^q}{\left(2q+1\right)!!} \ab{E_{\textbf{p}}^{n-2q} \left( p \cdot \Delta \cdot p\right)^q}_0^{(j)}, \label{I_nq:j}\\
        J_{nq}^{(j)} &\equiv \frac{\left(-1\right)^q}{\left(2q+1\right)!!} \ab{E_{\textbf{p}}^{n-2q} \left(p \cdot \Delta \cdot p \right)^q}_1^{(j)} \label{J_nq:j}.
    \end{align}
\end{subequations}
We can solve the counter term, $\phi_{\textbf{p}}^{(j)*}$ with the help of frame and matching conditions, which in the present case is assumed to be given by,
\begin{align}
    \sum_j \ab{g_{\textbf{p}} \phi^{(j)}_{1\textbf{p}}}_1^{(j)} = 0,
    \hspace{2cm}
    \sum_j \ab{q_{\textbf{p}} p^{\ab{\mu}} \phi^{(j)}_{1\textbf{p}}}_1^{(j)} = 0. \label{MC,FC-ERTA:j}
\end{align}
Using Eqs.~\eqref{phi^*_1:j}-\eqref{A_p^j-def} and defining $\delta T \equiv \mathcal{C}_2 \theta$, we obtain\footnote{The subscript `2' in $\mathcal{C}_2$ is used to be consistent with earlier works \cite{Dash:2020zqx, Bhadury:2024ckc}.},
\begin{align}
    \delta u^\mu = 0,
    \hspace{2cm}
    \mathcal{C}_2 \equiv \frac{\sum_{j} \ab{\left(\tau_{\rm R}/E_{\textbf{p}}\right) g_{\textbf{p}} A_{\textbf{p}}^{(j)}}_{1}^{(j)}}{\beta^2 \sum_{j} \ab{g_{\textbf{p}} E_{\textbf{p}}}_{1}^{(j)}}, \label{C_2-def:j}
\end{align}
where we used the fact that due to the normalization of fluid four-velocity, $u\cdot \delta u = \mathcal{O} \left(\partial^2\right)$. We are now ready to evaluate the solution, $\phi_{\textbf{p}}^{(j)}$, For this, we consider the expansion,
\begin{align}
    \phi_{\textbf{p}}^{(j)} = \sum_{n,l = 0}^\infty \Phi_{j,n}^{\ab{\mu_1\cdots\mu_l}} p_{\langle\mu_1} \cdots p_{\mu_l\rangle} P_{n}^{(j,l)}. \label{phi^ERTA-expansion:j}
\end{align}
where, $P_n^{(j,l)}$ is the $n$-th order orthogonal polynomial associated with the $l$-rank irreducible tensor for $j$-th species. These polynomials satisfy the orthogonality condition,
\begin{align}
    \frac{l!}{\left(2 l + 1\right)!!} \ab{\left(E_{\textbf{p}}/\tau_{\rm R}\right) \left(p \cdot \Delta \cdot p\right)^l P_n^{(j,l)} P_q^{(j,l)}}_1^{(j)} = A^{(j,l)}_{n} \delta_{nq}. \label{P_n-orthogonality:j}
\end{align}
where the expression of $A_{n}^{(j,l)}$ can be obtained by substituting $q\to n$. As in the case of single-species in Section \ref{sec:SSQ}, we define, $P_0^{(j,0)} = \beta E_{\textbf{p}}$ and, $P_0^{(j,l\geq1)} = 1$ for all `$N$' types of species. Substituting $\phi_{\textbf{p}}^{(j)}$ from Eq.~\eqref{phi^ERTA-expansion:j} in Boltzmann equation of $j$-th species in Eq.~\eqref{Beq-ERTA_j2} and taking moments with appropriate weights of the form $P_q^{(j,l)} p^{\langle\nu_1} \cdots p^{\nu_l\rangle}$ along with the orthogonality property of the polynomials, $P_q^{(j,l)}$ we can write,
\begin{subequations}
    \begin{align}
        \Phi_{j,n} &= \frac{1}{A_{j,n}^{(0)}} \left[\beta^2\, \mathcal{C}_2 \ab{\left(E_{\textbf{p}}^2/\tau_{\rm R}\right) P_{n}^{(j,0)}}_{1}^{(j)} - \ab{A^{(j)}_{\textbf{p}} P_{n}^{(j,0)}}_{1}^{(j)}\right] \theta \hspace{1.6cm} {\rm for,~} n\geq0. \label{Phi_j,n} \\
        \Phi_{j,n}^{\ab{\mu_1}} &= 0 \hspace{9.4cm} {\rm for,~} n\geq0, \label{Phi_j,n^m1} \\
        \Phi_{j,n}^{\ab{\mu_1\mu_2}} &= \frac{\beta \sigma^{\mu_1\mu_2} }{A_{n}^{(j,2)}} \ab{\left(2/15\right) P_{n}^{(j,2)} \left(p\cdot\Delta\cdot p\right)^2}_{1}^{(j)} \hspace{4cm} {\rm for,~} n\geq0, \label{Phi_j,n^m1m2} \\
        \Phi_{j,n}^{\ab{\mu_1\cdots\mu_{l}}} &= 0. \hspace{9.3cm} {\rm for,~} n\geq0, l\geq3 \label{Phi_j,n^m1..ml}
    \end{align}
\end{subequations}
One may further notice that, due to the relation, $\ab{A_{\textbf{p}}^{(j)} P_{n}^{(j,0)}}_1^{(j)} = 0$, it is possible to prove, $\Phi_{(j),0} P_{0}^{(j,0)} = \phi_{1\textbf{p}}^{*(j)}$. Similarly to Eq.~\eqref{G_p} we can write for the $j$-th species,
\begin{align}
    \mathcal{G}_{\textbf{p}}^{(j)} = \sum_{n=0}^\infty \frac{1}{A_n^{(j,l)}} \frac{l!}{\left(2 l + 1\right)!!} \ab{\left(E_{\textbf{p}}/\tau_{\rm R}\right) \mathcal{G}_{\textbf{p}}^{(j)} P_{n}^{(j,l)} \left(p\cdot\Delta\cdot p\right)^l}_1 P_n^{(j,l)}, \label{G_p:j}
\end{align}
allowing us to re-write, $\phi_{\textbf{p}}^{(j)}$ by substituting Eqs.~\eqref{Phi_j,n}-\eqref{Phi_j,n^m1..ml} in Eq.~\eqref{phi^ERTA-expansion:j} as,
\begin{align}
    \phi_{\textbf{p}}^{(j)} = \beta^2 E_{\textbf{p}} \delta T - \left(\frac{\tau_{\rm R} A_{\textbf{p}}^{(j)}}{E_{\textbf{p}}}\right) \theta + \left(\frac{\beta \tau_{\rm R}}{E_{\textbf{p}}}\right) p_{\langle\alpha} p_{\beta\rangle}\, \sigma^{\alpha\beta}. \label{phi_p-erta:j}
\end{align}

\subsection{Transport Coefficients}
\label{ssec:TC_j}

Considering the matching condition, $g_{\textbf{p}} = 1$, and using Eq.~\eqref{phi_p-erta:j} in Eqs.~\eqref{dele,delP-def_j} and \eqref{h^m,p^mn-def_j} along with Eq.~\eqref{NS-Eqs} we can write down the transport coefficients as,
\begin{subequations}
    \begin{align}
        e &= \sum_{j=1}^{N} \left[\ab{\left(\tau_{\rm R}/E_{\textbf{p}}\right) A_{\textbf{p}}^{(j)}}_{1}^{(j)} \left(\frac{\sum_{j=1}^{N} J_{30}^{(j)}}{\sum_{j=1}^{N} J_{10}^{(j)}}\right) - \ab{\tau_{\rm R} E_{\textbf{p}} A^{(j)}_{\textbf{p}}}_{1}^{(j)}\right], \label{e:j}\\
        \rho &= - \sum_{j=1}^{N} \left[\ab{\left(\tau_{\rm R}/E_{\textbf{p}}\right) A_{\textbf{p}}^{(j)}}_{1}^{(j)} \left(\frac{\sum_{j=1}^{N} J_{31}^{(j)}}{\sum_{j=1}^{N} J_{10}^{(j)}}\right) - \ab{\left(1/3\right) \left(\tau_{\rm R}/E_{\textbf{p}}\right) \left(p\cdot\Delta\cdot p\right) A^{(j)}_{\textbf{p}}}_{1}^{(j)}\right], \label{rho:j}\\
        \kappa &= 0, \label{kappa:j}\\
        %
        \eta &= \frac{\beta}{15} \sum_{j=1}^{N} \ab{\left(p\cdot\Delta\cdot p\right)^2 \left(\tau_{\rm R}/E_{\textbf{p}}\right)}_{1}^{(j)}. \label{eta:j}
    \end{align}
\end{subequations}
The frame invariant transport coefficients are determined through the entropy production, which for $N$-particle species is given by,
\begin{subequations}
    \begin{align}
        &\partial_\mu S^\mu_1 = - \sum_{j=1}^{N} \int_p C[f_{\textbf{p}}^{(j)}] \phi^{(j)}_{\textbf{p}}, \label{del.S1:j}\\
        &= \sum_{j=1}^N \!\left[\sum_{n=1}^\infty \Phi_{j,n}^2 A_n^{(j,0)} \!+\!\! \sum_{n=0}^\infty \Phi_{j,n}^{\ab{\mu}} \Phi_{n,\,\ab{\mu}} A_n^{(j,1)} \!+\!\! \sum_{n=0}^\infty \Phi_{j,n}^{\langle\mu_1 \mu_2\rangle} \Phi_{j,n,\,\langle\mu_1 \mu_{2}\rangle} A_n^{(j,2)} \!+\!\! \sum_{l=3}^\infty \sum_{n=0}^\infty \Phi_{j,n}^{\langle\mu_1 \cdots\mu_l\rangle} \Phi_{j,n,\,\langle\mu_1 \cdots \mu_{l}\rangle} A_n^{(j,l)}\right]\!. \label{del.S2:j}
    \end{align}
\end{subequations}
Substituting the expressions of $\Phi_{j,n}^{\ab{\mu_1\cdots\mu_l}}$ from Eqs.~\eqref{Phi_j,n}-\eqref{Phi_j,n^m1..ml} and comparing the results with Eq.~\eqref{del.S-final} we find the frame-invariant transport coefficients are given by,
\begin{subequations}
    \begin{align}
        \zeta_s &= \frac{1}{\beta} \sum_{j=1}^{N} \ab{\frac{A_{\textbf{p}}^{(j)} A_{\textbf{p}}^{(j)} \tau_{\rm R}}{E_{\textbf{p}}}}_1^{(j)}, \label{zeta_s:j}\\
        \eta_s &= \sum_{j=1}^{N} \frac{\beta}{15} \ab{\left(\tau_{\rm R}/E_{\textbf{p}}\right) \left(p\cdot\Delta\cdot p\right)^2}_1^{(j)}. \label{eta_s:j}
    \end{align}
\end{subequations}
In the next section, we will explore the temperature dependence of the obtained transport coefficients numerically. 

\section{Results and Discussions}
\label{sec:R&D}
To analyze the temperature variation of the shear and bulk viscous coefficients obtained in Eqs.~\eqref{zeta_s:j} and \eqref{eta_s:j}, we require two ingredients, (i) the dispersion relation for the quasi-particle species and (ii) the specific functional form of the relaxation time. We consider the on-shell dispersion relation, $E_{\textbf{p}_j}=\sqrt{\textbf{p}_j^2+m^2_j(T)}$ for the quasi-particles with temperature dependent effective mass given by 
    \begin{align}
        m_j(T)&=\sqrt{(m_j^{\rm B})^2+\Pi^{\rm int}_j(T)},
    \end{align}
where, the temperature independent constant $m_j^{\rm B}$ is the bare mass of the quasi-particle and $\Pi^{\rm int}_j(T)$ effectively incorporates  the interactions. For the description of the strongly  interacting 2+1 ($=N_f$) flavor hot QCD matter we consider non-degenerate strange  and light (up and down) quark-anti-quark sectors with bare masses $m_{s,\bar{s}}^{\rm B}=0.150$ GeV and $m_{u,\bar{u}}^{\rm B}=m_{d,\bar{d}}^{\rm B}=m_s^{\rm B}/28.15$ respectively along with  massless gluons. The interaction part is motivated from the asymptotic form of the Hard Thermal Loop (HTL) self-energy \cite{Bellac:2011kqa}  which is given by \cite{Mykhaylova:2020pfk}
    \begin{align}
       \Pi^{\rm int}_j(T)&=2\Big[m_j^{\rm B} \sqrt{\frac{g^2_{\rm eff}(T)}{6}T^2}+\frac{g^2_{\rm eff}(T)}{6}T^2\Big],
    \end{align}
for $j=\{u,d,s\}$ and their anti-particles whereas for gluons, 
\begin{align}
       \Pi^{\rm int}_g(T)&=\Big(N_c+\frac{N_f}{2}\Big)\frac{g^2_{\rm eff}(T)}{6}T^2,
    \end{align}
with number of colors, $N_c=3$. The temperature dependence of the effective coupling can be fixed  from the 2+1 flavor QCD  thermodynamics obtained in lattice calculation following the procedure outlined in Ref.~\cite{Rocha:2022fqz}. For this purpose, we consider the parametrized form of the trace anomaly  given by \cite{Nopoush:2015yga,Borsanyi:2010cj}
\begin{align}
    \frac{I(T)}{T^4}&=\Bigg(\frac{h_0}{1+h_3t^2}+f_0\frac{\tanh{(f_1t+f_2)}+1}{1+g_1t+g_2t^2}\Bigg)\exp[-(h_1/t)-(h_2/t^2)],
    \label{trace_anomaly}
\end{align}
 with $t=T/(0.2~{\rm GeV})$ and we consider the following parameter set for $N_f=2+1$:
 \begin{subequations}
     \begin{align}
       (h_0,h_1,h_2,h_3)&=(0.1396,-0.18,0.035,0.01),\\(f_0,f_1,f_2)&=(2.76,6.79,-5.29),\\ (g_1,g_2)&=(-0.47,1.04).   
     \end{align}
 \end{subequations}

\begin{figure*}[t!]
    \centering
       \begin{subfigure}[t]{0.49\linewidth}
        \centering
         \includegraphics[width=\linewidth]{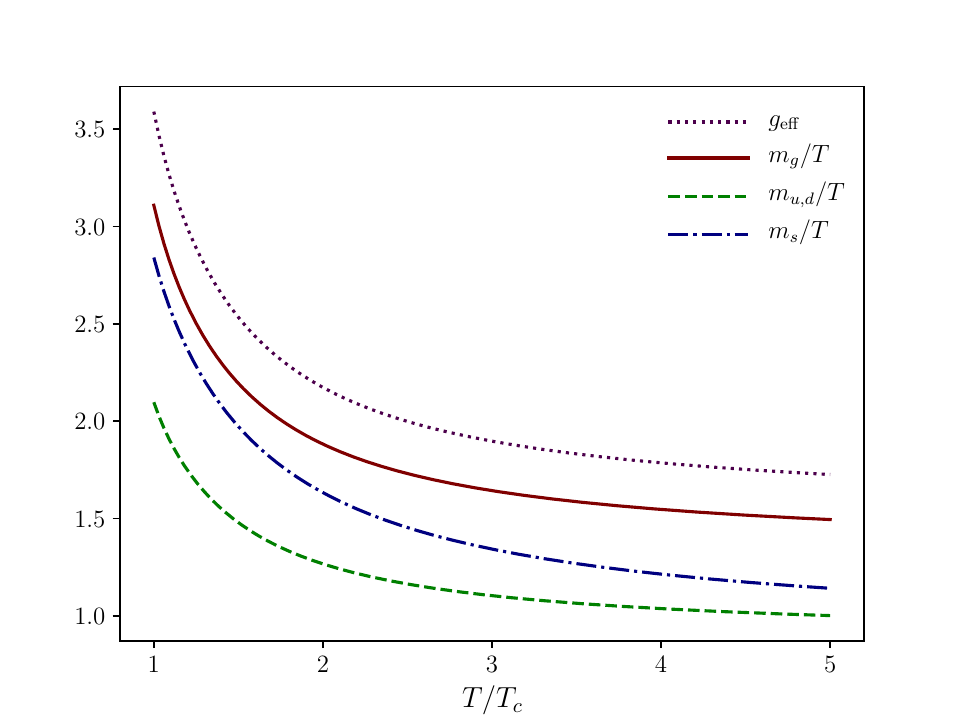}
        \caption{}
        \label{massga}
    \end{subfigure}%
    ~ 
    \begin{subfigure}[t]{0.49\linewidth}
        \centering
        \includegraphics[width=\linewidth]{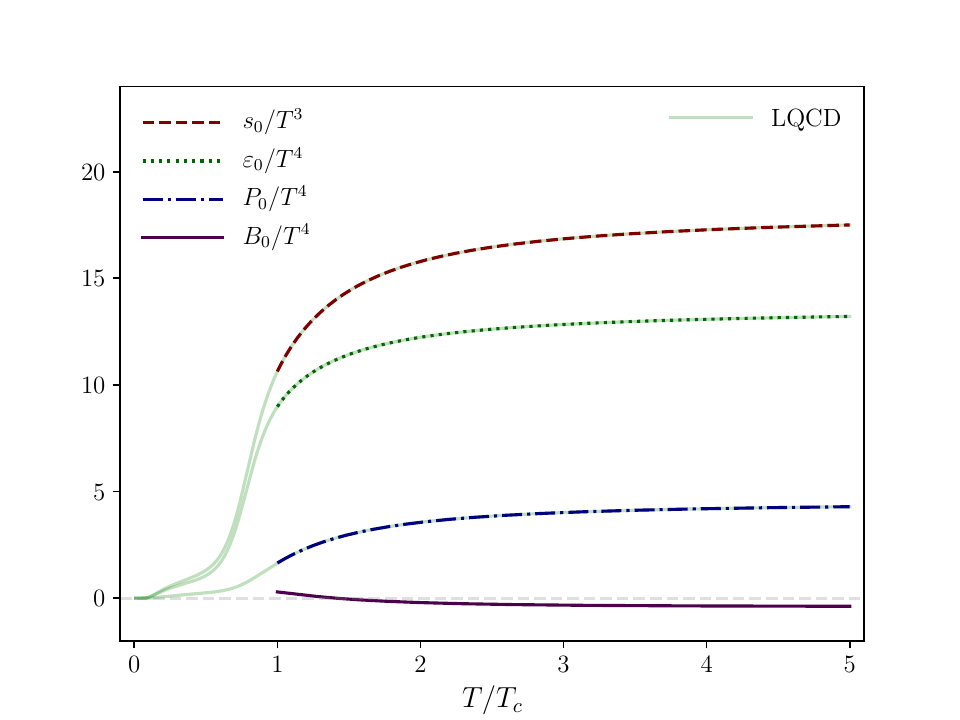}
        \caption{}
        \label{massgb}
    \end{subfigure}
    \caption{
            \justifying
            \noindent
             (a) Variation of the effective coupling (purple dotted) and scaled quasi-particle masses ($m_g/T$, $m_{u,d}/T$ and $m_s/T$) are shown (in red solid, green dashed and blue dot-dashed style respectively) as a function of $T/T_c$ where $T_c=0.155$ GeV. (b) The equilibrium thermodynamic quantities obtained in the quasi-particle approach, namely the scaled entropy density ($s_0/T^3$), the scaled energy density ($\varepsilon_0/T^4$), and the scaled pressure ($P_0/T^4$) are shown (in red dashed, green dotted and blue dot-dashed style) as a function of $T/T_c$. The corresponding LQCD results (light-green solid) obtained from the parametrization of the trace anomaly are also plotted for comparison. The solid-purple curve corresponds to the variation of the equilibrium bag contribution ($B_0/T^4$) in the quasi-particle approach.
            }
\end{figure*}

 Considering the pseudo-critical temperature $T_c=0.155$ GeV, the  extracted effective coupling is shown in Fig. \ref{massga} as a function of the scaled variable $T/T_c$ in the range $1\le T/T_c\le5$. As the temperature is increased, a rapid decrease in the effective coupling near $T_c$ is observed which eventually almost saturates at higher temperature. This is similar to the observations obtained in quasi-particle models where a parametrization of the effective coupling is considered based on the functional form of one and two loop QCD running coupling \cite{Plumari:2011mk,Bluhm:2007nu}. This trend in effective coupling is also reflected in the temperature variation of the scaled effective masses of the gluon ($m_g/T)$, light ($m_{u,d}/T)$ and strange ($m_s/T)$  quark sectors as  shown in 
 Fig.~\ref{massga}. In fact, the bare mass $m_g^{\rm B}$ being zero, the temperature variation of  the scaled effective mass for gluon quasi-particle becomes identical to that of the effective coupling up to an overall constant factor $\sqrt{3/4}$. Similar reasoning also applies to up and down quarks with small $m_{u,d}^{\rm B}$ (with an overall factor around $1/\sqrt{3}$). However, the bare mass term does play a role for the strange quarks resulting in a qualitatively different temperature profile in comparison to the lighter counterparts. Also note that the  hierarchy of the effective masses $m_g\ge m_s \ge m_{u,d}$ throughout the considered temperature range  is consistent with other quasi-particle approaches with $N_f=2+1$ \cite{Mykhaylova:2019wci,Mykhaylova:2020pfk,Ma:2018bwf}. The equilibrium thermodynamic quantities of the QCD medium, namely the scaled entropy density ($s_0/T^3$), the scaled energy density ($\varepsilon_0/T^4$), and the scaled pressure ($P_0/T^4$) from the lattice parametrization are shown in Fig.~\ref{massgb} and compared with those obtained in the quasi-particle model with  temperature dependent effective masses. As can be seen from the figure, the energy density and pressure defined in Eq.\eqref{e0,P0-def_j} with the bag contribution ($B_0$) reproduces the results obtained from the parametrized  interaction measure (given in Eq.~\eqref{trace_anomaly}) within the considered range of temperature $1\le T/T_c\le5$. The temperature variation of the scaled bag contribution $B_0/T^4$ is also shown in Fig.~\ref{massgb}. As can be noticed, the bag contribution (apart from a tiny interval close to $T_c$) is  negative and it remains an order of magnitude smaller compared to the energy density and pressure throughout the considered temperature range. This is consistent with the results obtained in Ref.~\cite{Rocha:2022fqz}.                

For the momentum dependence, we restrict ourselves to the simple power law ansatz  which in the fluid rest frame is given by $\tau_{\rm R} = t_{\rm R} \left(\beta E_{\textbf{p}}\right)^\ell$. We explore the influence of the exponent in the range $0\le\ell\le1$. The lower limit describes the momentum independent case whereas the $\ell=1$ corresponds to the scalar field theory \cite{Denicol:2022bsq}. Note that $\ell=0.5$ is commonly quoted as the value of the exponent for QCD effective kinetic theory and it lies in the midway between the lower and upper limit of the power law exponent considered here. 

\begin{figure*}[t!]
    \centering
       \begin{subfigure}[t]{0.49\linewidth}
        \centering
         \includegraphics[width=\linewidth]{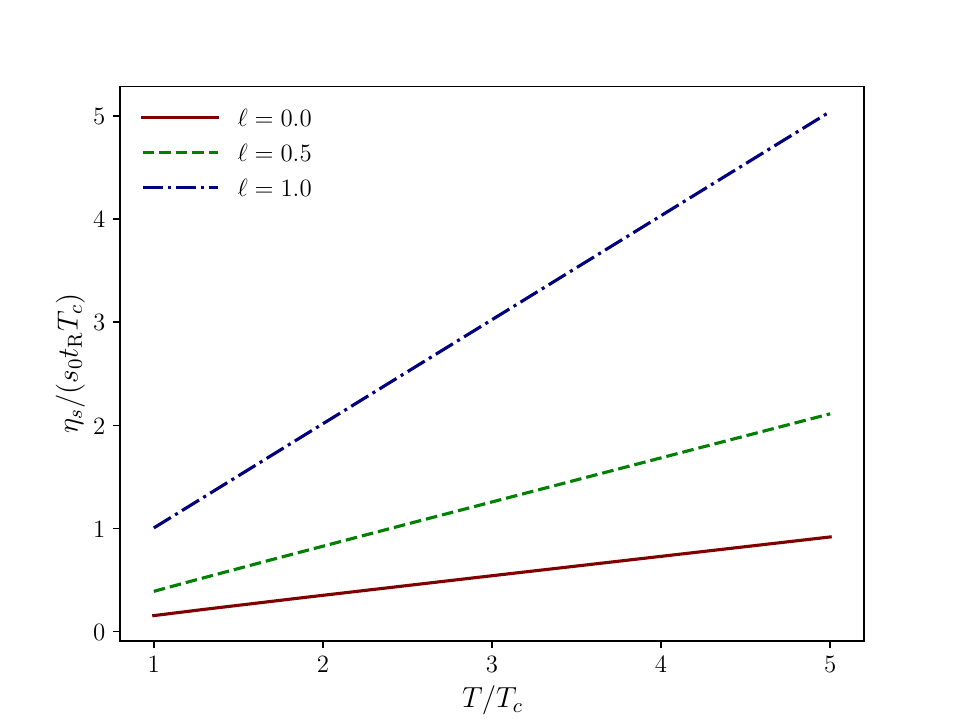}
        \caption{}
        \label{etabysa}
    \end{subfigure}%
    ~ 
    \begin{subfigure}[t]{0.49\linewidth}
        \centering
       \includegraphics[width=\linewidth]{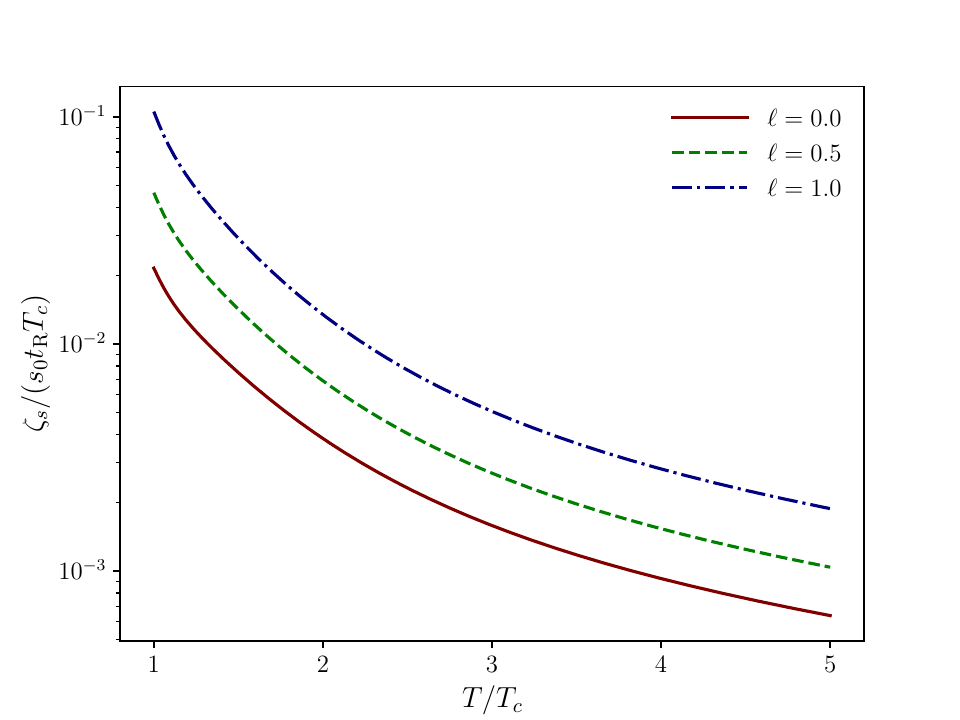}
        \caption{}
        \label{zetabysb}
    \end{subfigure}
    \caption{
            \justifying
            \noindent
            (a) Variation of the ratio of the frame independent shear viscosity coefficient ($\eta_s$ scaled by $t_R T_c$) and equilibrium entropy density ($s_0$) is shown (in red solid, green dashed and blue dot-dashed style respectively) as a function of $T/T_c$ (where $T_c=0.155$ GeV) for fixed set of values of the power law exponent $\ell=(0.0,0.5,1.0)$. (b) Variation of the ratio of the frame independent bulk viscosity coefficient ($\zeta_s$ scaled by $t_R T_c$) and equilibrium entropy density ($s_0$) is shown (in red solid, green dashed and blue dot-dashed style respectively) as a function of $T/T_c$ (where $T_c=0.155$ GeV) for fixed set of values of the power law exponent $\ell=(0.0,0.5,1.0)$.
            }
\end{figure*}

Apart from the power law exponent, the functional form of the relaxation time also depends on the parameter $t_{\rm R}$ which though momentum independent can be a function of temperature. As our primary objective is to analyze the influence of the power law exponent in the QCD transport, in the present work we do not attempt to obtain  the temperature dependence of $t_{\rm R}$ explicitly. Rather, we scale the  transport coefficients by an overall  dimensionless factor $t_{\rm R} T_c$ and study the effect of the power law exponent on  the residual temperature dependence. The temperature variation of the scaled shear ($\eta_s/s_0 t_{\rm R} T_c$) and bulk ($\zeta_s/s_0 t_{\rm R} T_c$) viscous coefficients are shown in Figs.~\ref{etabysa} and \ref{zetabysb} respectively for fixed set of values of the power law exponent $\ell=(0.0,0.5,1.0)$. It can be observed that as the temperature increases from $T_c$,  $\eta_s/s_0 t_{\rm R} T_c$  increases whereas $\zeta_s/s_0 t_{\rm R} T_c$ shows a decreasing trend for all the $\ell$  values considered. Moreover, at a fixed temperature, with larger $\ell$  the  specific shear coefficient increases whereas the  specific bulk coefficient decreases. This opposite trend of shear and bulk for different values of the power law exponent is consistent with the transport coefficients obtained  with single quasi-particle species \cite{Rocha:2022fqz}.

\begin{figure*}[t!]
    \centering
       \begin{subfigure}[t]{0.49\linewidth}
        \centering
         \includegraphics[width=\linewidth]{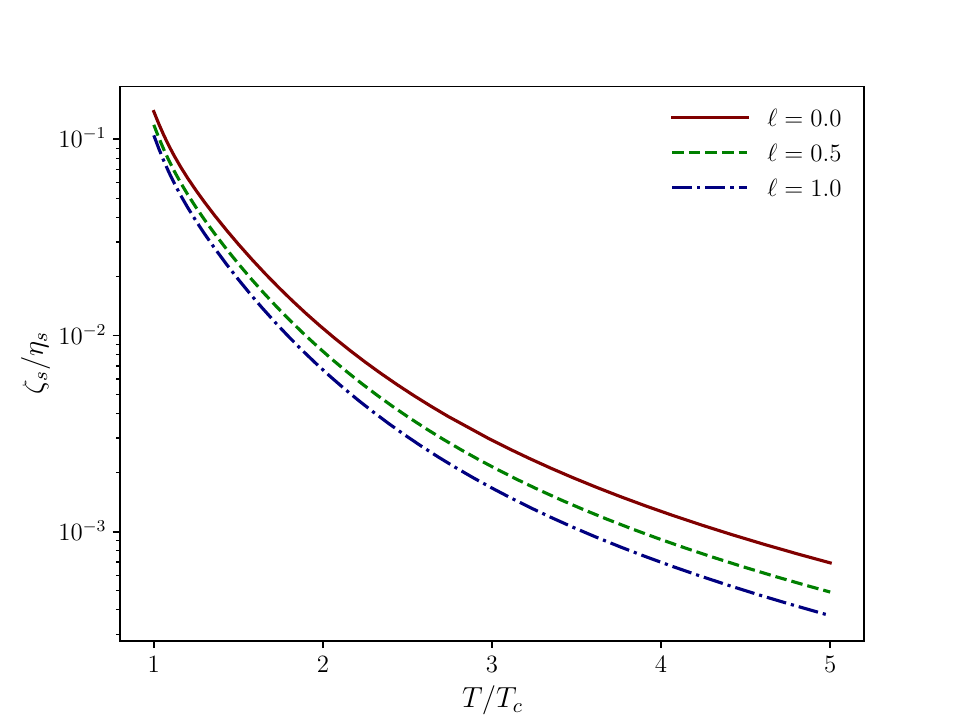}
        \caption{}
        \label{zetabetaT}
    \end{subfigure}%
    ~ 
    \begin{subfigure}[t]{0.49\linewidth}
        \centering
       \includegraphics[width=\linewidth]{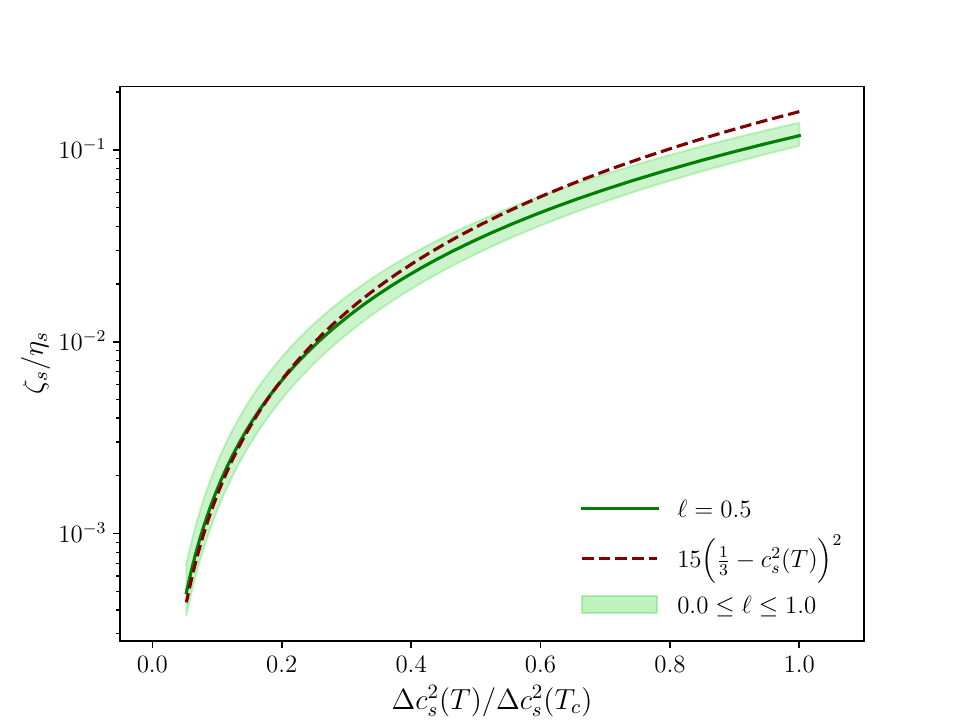}
        \caption{}
        \label{zetabetacon}
    \end{subfigure}
    \caption{
            \justifying
            \noindent
             (a) Variation of the ratio of the frame independent transport coefficients ($\zeta_s/\eta_s$) is shown (in red solid, green dashed and blue dot-dashed style respectively) as a function of $T/T_c$ (where $T_c=0.155$ GeV) for fixed set of values of the power law exponent $\ell=(0.0,0.5,1.0)$. (b) Variation of the ratio of the frame independent transport coefficients ($\zeta_s/\eta_s$) is shown  as a function of  $\Delta c_s^2(T)=\frac{1}{3}-c_s^2(T)$ (scaled by $\Delta c_s^2(T_c)$). The green patch corresponds to the $\zeta_s/\eta_s$ values for  $0.0\le\ell\le1.0$. The  variation corresponding to $\ell=0.5$ (shown in green solid style) and the case  $\zeta_s/\eta_s=15\big(\frac{1}{3}-c_s^2(T)\big)^2$ (shown in red dashed style) are also plotted for comparison.
    }
    \end{figure*}

The temperature dependence of the ratio of the frame independent transport coefficients ($\zeta_s/\eta_s$) is shown in Fig.~\ref{zetabetaT} for  $\ell=(0.0,0.5,1.0)$.  Apart from the overall decreasing trend, it can be noticed  that  higher values of the power law exponent results in  a reduced $\zeta_s/\eta_s$ throughout the  temperature range. Also note that  this reduction in the $\zeta_s/\eta_s$ ratio for $\ell>0.0$  is more prominent at higher values of temperature compared to the region close to $T_c$.   It should be mentioned here  that for weakly coupled QCD, a scaling law exists for the bulk to shear ratio which is given by \cite{Arnold:2006fz,Arnold:2000dr,Arnold:2003zc}
\begin{align}
   \zeta_s/\eta_s=15\Big(\frac{1}{3}-c_s^2(T)\Big)^2,
   \label{weakscaling}
\end{align}
which is in fact similar to the estimation for a gas of photon first obtained in \cite{Weinberg:1971mx}. On the other hand, in the  strongly coupled scenario,  $\zeta_s/\eta_s\sim\Big(\frac{1}{3}-c_s^2(T)\Big)$ \cite{Buchel:2005cv} (see Refs.~\cite{Buchel:2007mf, Schafer:2009dj} for a detailed discussion on transport properties in the strongly coupled regime). As already discussed, in the present formulation the ratio of the frame independent transport coefficients depends on the power law exponent parameter $\ell$.  Thus it is interesting to analyze the dependence of $\zeta_s/\eta_s$ on $\ell$  as the conformal measure $\Delta c_s^2(T)=\frac{1}{3}-c_s^2(T)$ varies with temperature. In Fig.~\ref{zetabetacon}, the  variation of $\zeta_s/\eta_s$  is shown as a function of $\Delta c_s^2(T)$ (scaled by $\Delta c_s^2(T_c)$) for $0.0\le\ell\le1.0$. The ratio increases with the scaled conformal measure and it can be observed that it is more suppressed  for higher $\ell$ values compared to the momentum independent scenario $\ell=0.0$.  $\zeta_s/\eta_s$ for $\ell=0.5$ as well as the scaling relation of the weakly coupled QCD  given in Eq.\eqref{weakscaling} are also plotted in Fig.~\ref{zetabetacon} for comparison. It can be seen that at higher temperature (close to the origin in the $\Delta c_s^2(T)$ axis), the ratio in the quasi-particle formulation with $\ell=0.5$ is similar to the estimate from the weakly coupled scenario. However we find that close to $T=T_c$, the weakly coupled approach overestimates the  ratio. 
\begin{figure*}[t!]
    \centering
    \includegraphics[width=0.49\linewidth]{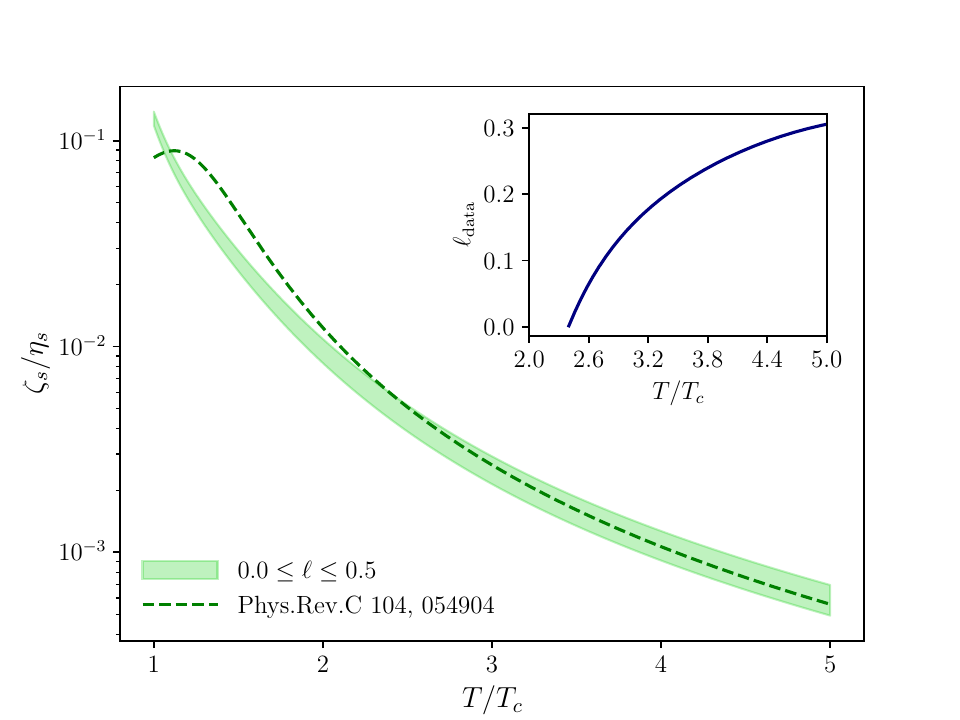}
        \caption{
            \justifying
            \noindent
             The best-fit parametrization  $\big(\zeta_s/\eta_s\big)_{\rm data}$ obtained in Ref.~\cite{Parkkila:2021tqq} is shown (in green dashed style) as a function of $T/T_c$. The light green band corresponds to the $\zeta_s/\eta_s$ values in the quasi-particle model for $0.0\le\ell\le0.5$. The temperature dependence of the  power law exponent extracted from the best-fit parametrization is shown in the inset.
                         }
    \label{fig:zeta/eta-jetscape}
    \end{figure*}

So far, we have analyzed the influence of the power law exponent on the transport coefficient considering a fixed set of $\ell$ values  in the range $[0,1]$. However, to achieve a better understanding of the applicability of the the quasi-particle model as formulated in the present study, it is desirable to compare the  results with realistic scenarios. For this purpose, let us focus on the  results obtained in the  data driven approach based on the Bayesian analysis \cite{Bernhard:2019bmu}. In Ref.~\cite{Bernhard:2019bmu}, the temperature dependence of the specific shear viscosity  has been parametrized by a three parameter linear ansatz  whereas a three-parameter unnormalized Cauchy distribution is used to  parametrize the temperature dependence of  the specific bulk viscosity. Further improvement of the parameterization is obtained in Ref.~\cite{Parkkila:2021tqq} utilizing larger statistics as well as additional observables. In the present work we consider the best-fit values obtained in \cite{Parkkila:2021tqq} to parametrize the frame independent  bulk to shear ratio as 
\begin{align}
    \big(\zeta_s/\eta_s\big)_{\rm data}=\frac{(\zeta_s/s_0)_{\rm max}}{\Big[1+\Big(\frac{T-(\zeta_s/s_0)_{\rm peak}}{(\zeta_s/s_0)_{\rm width}}\Big)^2\Big]\Big[(\eta_s/s_0)(\tilde{T}_c)+(\eta_s/s_0)_{\rm slope}(T-\tilde{T}_c)\Big(\frac{T}{\tilde{T}_c}\Big)^{(\eta_s/s_0)_{\rm crv}}\Big]},
    \label{bayesian}
\end{align}
with $\tilde{T}_c=0.147$ GeV and 
\begin{subequations}
     \begin{align}
       ((\zeta_s/s_0)_{\rm max},(\zeta_s/s_0)_{\rm peak},(\zeta_s/s_0)_{\rm width})&=(0.010,0.170,0.057),\\((\eta_s/s_0)(\tilde{T}_c),(\eta_s/s_0)_{\rm slope},(\eta_s/s_0)_{\rm crv})&=(0.104,0.425,-0.738).   
     \end{align}
 \end{subequations}
In Fig.~\ref{fig:zeta/eta-jetscape}, we compare the above parametrization  with the $\zeta_s/\eta_s$ obtained from the quasi-particle approach with fixed values of the power law exponent in the range $0.0\le\ell\le0.5$ shown in light green band. It can be noticed that beyond $T\sim2.5T_c$, the best-fit parametrization from the Bayesian analysis lies well within the $0.0\le\ell\le0.5$ band and in fact never reaches the edge corresponding to $\ell=0.5$  up to $T=5T_c$. One can extract the power law exponent corresponding to the  best-fit parametrization (say $\ell=\ell_{\rm data}$)  in the overlap region by  equating the ratio in the two scenarios as
\begin{align}
    \big(\zeta_s/\eta_s\big)_{\ell=\ell_{\rm data}}=\big(\zeta_s/\eta_s\big)_{\rm data},
\end{align}
for a given temperature. In Fig.~\ref{fig:zeta/eta-jetscape}, the temperature dependence of the extracted $\ell_{\rm data}(T)$ is shown in the inset. It can be noticed that the power law exponent smoothly increases with temperature reaching up to $\ell=0.3$ at $T=5T_c$. Also, it should be noted that at temperatures close to $T_c$, negative values of the power law exponent is required to match the best-fit result.

\section{Conclusion and Outlook}
\label{sec:C&O}
In this work, we formulate the first order relativistic dissipative hydrodynamics of a system of quasi-particles from kinetic theory with modified relaxation time approximation. The primary objective being the incorporation of the momentum dependence of the relaxation time, we consider two approaches based on the modified collision kernels (i) NRTA and (ii) ERTA. For single quasi-particle species, we have shown that both approaches are consistent with each other providing the same out-of-equilibrium corrections to phase-space distribution and hence the transport coefficients. We generalize the extended relaxation time approach to incorporate multiple quasi-particle species and obtain the frame and matching invariant transport coefficients from the entropy production. The formalism is then implemented to study the shear and bulk viscosity coefficients of the strongly interacting 2+1  flavor hot QCD medium with quasi-gluon and (light and strange) quasi-quark (anti-quark) sectors. The quasi-particle masses are temperature-dependent due to the interaction term present in the dispersion relation. In the present study we consider the interaction term with temperature dependent effective coupling $g_{\rm eff}(T)$. The functional form of the interaction term is motivated from the asymptotic form of the HTL self-energy. However, in the present study, we do not consider the perturbative running of the coupling. Rather the temperature dependence of the effective coupling is fixed using the parametrized form of the trace anomaly obtained in the $2+1$ flavour LQCD simulation \cite{Borsanyi:2010cj} and the extracted $g_{\rm eff}(T)$ essentially fixes the temperature dependence of the equilibrium thermodynamics. 

 To study the transport coefficient of the hot QCD medium, we consider the power law ansatz $\tau_{\rm R} = t_{\rm R} \left(\beta E_{\textbf{p}}\right)^\ell$ for the momentum dependence of the relaxation time. We focus on the influence of the power law exponent $\ell$  on the frame choice and matching condition independent shear and bulk viscosity coefficient and also on the bulk to shear ratio $\zeta_s/\eta_s$ where the momentum independent part $t_{\rm R}$ cancels. For the power law exponent, we consider the range of values $0.0\le\ell\le1.0$ and find that specific shear viscosity coefficient (scaled by $t_{\rm R} T_c$) significantly increases for higher $\ell$ compared to the $\ell=0.0$ case whereas the specific bulk viscous coefficient (scaled by $t_{\rm R} T_c$) gets significantly reduced. As a consequence, an overall reduction in the ratio $\zeta_s/\eta_s$ is observed for higher $\ell$ throughout the considered temperature range $1\le T/T_c\le5$. We also analyze the variation of $\zeta_s/\eta_s$ as a function of the conformal measure $\Delta c_s^2(T)$. We find that at higher temperature the estimation of the bulk to shear ratio for $\ell=0.5$ is well described by the scaling relation $\zeta_s/\eta_s=15[(1/3)-c_s^2(T)]^2$. However, significant deviation is observed close to $T_c$. Finally, we compare the $\zeta_s/\eta_s$  with the best-fit parametrization obtained in the data-driven Bayesian analysis \cite{Parkkila:2021tqq}. We find that for temperatures higher than  $T\sim2.5 T_c$, the parametrized result can be well described using the power law ansatz with a temperature-dependent power law exponent. The exponent extracted from the best-fit result shows a gradual increasing trend with temperature reaching up to $\ell=0.3$ at $T=5T_c$. This is less than the commonly quoted  value of the exponent for QCD which is $\ell=0.5$. 
 It should be mentioned that the present hydrodynamic formulation for the multiple quasi-particle species considers a single relaxation time as a simplification, which needs to be generalized for a realistic scenario. However, the assumption of a single relaxation time facilitates the extraction of the temperature dependence of the power law exponent simply from the ratio of transport coefficients. Moreover, the compatibility of the $\zeta_s/\eta_s$ results from the data-driven analysis and the quasi-particle scenario beyond $T\sim2.5 T_c$ is indeed promising and motivates to analyze the impact of the momentum-dependent relaxation time in the presence of non-zero chemical potential \cite{Daher:2024vxk}. It will also be interesting to implement this framework to study the dynamical evolution of the strongly interacting matter in the presence of background magnetic field \cite{Denicol:2018rbw, Panda:2020zhr}. Furthermore, looking into the growing interest in the spin physics of heavy-ion collisions, one may consider formulating a spin hydrodynamics \cite{Bhadury:2024ckc} for spin-1/2 quasi-particles with temperature-dependent effective mass. We leave such analysis for future exploration.

\section{Acknowledgment}

S. B. would like to thank Gabriel S. Rocha for some clarifying discussions. The authors acknowledge Amaresh Jaiswal for insightful correspondence. S. B. acknowledges the support of the Faculty of Physics, Astronomy, and Applied Computer Science, Jagiellonian University via Grant No. LM/36/BS. P. S. is funded by the EU’s NextGenerationEU instrument through the National Recovery and Resilience Plan of Romania - Pillar III-C9-I8, managed by the Ministry of Research, Innovation and Digitization, within the project entitled ``Facets of Rotating Quark-Gluon Plasma'' (FORQ), contract no. 760079/23.05.2023 code CF 103/15.11.2022.

\appendix

\section{\label{TIP}Thermodynamic Integrals and Properties of \texorpdfstring{$\delta$}{~}-function}

The thermodynamic integrals obey the following identities,
\begin{align}
    \Dot{I}_{nq} &= \Dot{\beta} \left[ \frac{1}{2} \frac{\partial\, m^2}{\partial \beta} \Big\{ (n-1) I_{n-2,q} - \beta J_{n-1,q} \Big\} - J_{n+1,q} \right] \label{DI_nq} \\
    X_{nq} &= \frac{1}{\left(2q+1\right)} \Big[ m^2 X_{n-2,q-1} - X_{n,q-1} \Big] \label{X_nq-RR}
\end{align}
where, $X_{nq}$ can be either of $I_{nq}$ or $J_{nq}$. In deriving Eq.~\eqref{DI_nq} we have used the relation,
\begin{align}
    D\delta(p^2 - m^2) &= - \frac{m (\partial_\mu m)}{(u\cdot p)} u^\mu D_{(p)} \delta(p^2 - m^2) \label{delta-property1}
\end{align}
where, $D \equiv (u\cdot\partial)$ and $D_{(p)} \equiv u\cdot\partial_{(p)}$. One can also show that,
\begin{align}
    \left(p \cdot \partial\right) \delta(p^2-m^2) &= - m (\partial_\mu m) \partial^\mu_{(p)}\delta(p^2 - m^2) \label{delta-property2}
\end{align}
Note that the relations in Eqs.~\eqref{DI_nq}-\eqref{delta-property2} remain true for any $j$-th particle species i.e. we can use the subscript $j$ in these equations and they will be satisfied.

\bibliography{ref}{}

\begin{thebibliography}{62}%
\makeatletter
\providecommand \@ifxundefined [1]{%
 \@ifx{#1\undefined}
}%
\providecommand \@ifnum [1]{%
 \ifnum #1\expandafter \@firstoftwo
 \else \expandafter \@secondoftwo
 \fi
}%
\providecommand \@ifx [1]{%
 \ifx #1\expandafter \@firstoftwo
 \else \expandafter \@secondoftwo
 \fi
}%
\providecommand \natexlab [1]{#1}%
\providecommand \enquote  [1]{``#1''}%
\providecommand \bibnamefont  [1]{#1}%
\providecommand \bibfnamefont [1]{#1}%
\providecommand \citenamefont [1]{#1}%
\providecommand \href@noop [0]{\@secondoftwo}%
\providecommand \href [0]{\begingroup \@sanitize@url \@href}%
\providecommand \@href[1]{\@@startlink{#1}\@@href}%
\providecommand \@@href[1]{\endgroup#1\@@endlink}%
\providecommand \@sanitize@url [0]{\catcode `\\12\catcode `\$12\catcode `\&12\catcode `\#12\catcode `\^12\catcode `\_12\catcode `\%12\relax}%
\providecommand \@@startlink[1]{}%
\providecommand \@@endlink[0]{}%
\providecommand \url  [0]{\begingroup\@sanitize@url \@url }%
\providecommand \@url [1]{\endgroup\@href {#1}{\urlprefix }}%
\providecommand \urlprefix  [0]{URL }%
\providecommand \Eprint [0]{\href }%
\providecommand \doibase [0]{http://dx.doi.org/}%
\providecommand \selectlanguage [0]{\@gobble}%
\providecommand \bibinfo  [0]{\@secondoftwo}%
\providecommand \bibfield  [0]{\@secondoftwo}%
\providecommand \translation [1]{[#1]}%
\providecommand \BibitemOpen [0]{}%
\providecommand \bibitemStop [0]{}%
\providecommand \bibitemNoStop [0]{.\EOS\space}%
\providecommand \EOS [0]{\spacefactor3000\relax}%
\providecommand \BibitemShut  [1]{\csname bibitem#1\endcsname}%
\let\auto@bib@innerbib\@empty
\bibitem [{\citenamefont {Ollitrault}(2008)}]{Ollitrault:2007du}%
  \BibitemOpen
  \bibfield  {author} {\bibinfo {author} {\bibfnamefont {J.-Y.}\ \bibnamefont {Ollitrault}},\ }\href {\doibase 10.1088/0143-0807/29/2/010} {\bibfield  {journal} {\bibinfo  {journal} {Eur. J. Phys.}\ }\textbf {\bibinfo {volume} {29}},\ \bibinfo {pages} {275} (\bibinfo {year} {2008})},\ \Eprint {http://arxiv.org/abs/0708.2433} {arXiv:0708.2433 [nucl-th]} \BibitemShut {NoStop}%
\bibitem [{\citenamefont {Florkowski}(2010)}]{Florkowski:2010zz}%
  \BibitemOpen
  \bibfield  {author} {\bibinfo {author} {\bibfnamefont {W.}~\bibnamefont {Florkowski}},\ }\href@noop {} {\emph {\bibinfo {title} {{Phenomenology of Ultra-Relativistic Heavy-Ion Collisions}}}}\ (\bibinfo {year} {2010})\BibitemShut {NoStop}%
\bibitem [{\citenamefont {Jaiswal}\ and\ \citenamefont {Roy}(2016)}]{Jaiswal:2016hex}%
  \BibitemOpen
  \bibfield  {author} {\bibinfo {author} {\bibfnamefont {A.}~\bibnamefont {Jaiswal}}\ and\ \bibinfo {author} {\bibfnamefont {V.}~\bibnamefont {Roy}},\ }\href {\doibase 10.1155/2016/9623034} {\bibfield  {journal} {\bibinfo  {journal} {Adv. High Energy Phys.}\ }\textbf {\bibinfo {volume} {2016}},\ \bibinfo {pages} {9623034} (\bibinfo {year} {2016})},\ \Eprint {http://arxiv.org/abs/1605.08694} {arXiv:1605.08694 [nucl-th]} \BibitemShut {NoStop}%
\bibitem [{\citenamefont {Florkowski}\ \emph {et~al.}(2018)\citenamefont {Florkowski}, \citenamefont {Heller},\ and\ \citenamefont {Spalinski}}]{Florkowski:2017olj}%
  \BibitemOpen
  \bibfield  {author} {\bibinfo {author} {\bibfnamefont {W.}~\bibnamefont {Florkowski}}, \bibinfo {author} {\bibfnamefont {M.~P.}\ \bibnamefont {Heller}}, \ and\ \bibinfo {author} {\bibfnamefont {M.}~\bibnamefont {Spalinski}},\ }\href {\doibase 10.1088/1361-6633/aaa091} {\bibfield  {journal} {\bibinfo  {journal} {Rept. Prog. Phys.}\ }\textbf {\bibinfo {volume} {81}},\ \bibinfo {pages} {046001} (\bibinfo {year} {2018})},\ \Eprint {http://arxiv.org/abs/1707.02282} {arXiv:1707.02282 [hep-ph]} \BibitemShut {NoStop}%
\bibitem [{\citenamefont {Romatschke}\ and\ \citenamefont {Romatschke}(2019)}]{Romatschke:2017ejr}%
  \BibitemOpen
  \bibfield  {author} {\bibinfo {author} {\bibfnamefont {P.}~\bibnamefont {Romatschke}}\ and\ \bibinfo {author} {\bibfnamefont {U.}~\bibnamefont {Romatschke}},\ }\href {\doibase 10.1017/9781108651998} {\emph {\bibinfo {title} {{Relativistic Fluid Dynamics In and Out of Equilibrium}}}},\ Cambridge Monographs on Mathematical Physics\ (\bibinfo  {publisher} {Cambridge University Press},\ \bibinfo {year} {2019})\ \Eprint {http://arxiv.org/abs/1712.05815} {arXiv:1712.05815 [nucl-th]} \BibitemShut {NoStop}%
\bibitem [{\citenamefont {Jeon}\ and\ \citenamefont {Heinz}(2015)}]{Jeon:2015dfa}%
  \BibitemOpen
  \bibfield  {author} {\bibinfo {author} {\bibfnamefont {S.}~\bibnamefont {Jeon}}\ and\ \bibinfo {author} {\bibfnamefont {U.}~\bibnamefont {Heinz}},\ }\href {\doibase 10.1142/S0218301315300106} {\bibfield  {journal} {\bibinfo  {journal} {Int. J. Mod. Phys. E}\ }\textbf {\bibinfo {volume} {24}},\ \bibinfo {pages} {1530010} (\bibinfo {year} {2015})},\ \Eprint {http://arxiv.org/abs/1503.03931} {arXiv:1503.03931 [hep-ph]} \BibitemShut {NoStop}%
\bibitem [{\citenamefont {Meyer}(2011)}]{Meyer:2011gj}%
  \BibitemOpen
  \bibfield  {author} {\bibinfo {author} {\bibfnamefont {H.~B.}\ \bibnamefont {Meyer}},\ }\href {\doibase 10.1140/epja/i2011-11086-3} {\bibfield  {journal} {\bibinfo  {journal} {Eur. Phys. J. A}\ }\textbf {\bibinfo {volume} {47}},\ \bibinfo {pages} {86} (\bibinfo {year} {2011})},\ \Eprint {http://arxiv.org/abs/1104.3708} {arXiv:1104.3708 [hep-lat]} \BibitemShut {NoStop}%
\bibitem [{\citenamefont {Bernhard}\ \emph {et~al.}(2019)\citenamefont {Bernhard}, \citenamefont {Moreland},\ and\ \citenamefont {Bass}}]{Bernhard:2019bmu}%
  \BibitemOpen
  \bibfield  {author} {\bibinfo {author} {\bibfnamefont {J.~E.}\ \bibnamefont {Bernhard}}, \bibinfo {author} {\bibfnamefont {J.~S.}\ \bibnamefont {Moreland}}, \ and\ \bibinfo {author} {\bibfnamefont {S.~A.}\ \bibnamefont {Bass}},\ }\href {\doibase 10.1038/s41567-019-0611-8} {\bibfield  {journal} {\bibinfo  {journal} {Nature Phys.}\ }\textbf {\bibinfo {volume} {15}},\ \bibinfo {pages} {1113} (\bibinfo {year} {2019})}\BibitemShut {NoStop}%
\bibitem [{\citenamefont {Nijs}\ \emph {et~al.}(2021)\citenamefont {Nijs}, \citenamefont {van~der Schee}, \citenamefont {G\"ursoy},\ and\ \citenamefont {Snellings}}]{Nijs:2020roc}%
  \BibitemOpen
  \bibfield  {author} {\bibinfo {author} {\bibfnamefont {G.}~\bibnamefont {Nijs}}, \bibinfo {author} {\bibfnamefont {W.}~\bibnamefont {van~der Schee}}, \bibinfo {author} {\bibfnamefont {U.}~\bibnamefont {G\"ursoy}}, \ and\ \bibinfo {author} {\bibfnamefont {R.}~\bibnamefont {Snellings}},\ }\href {\doibase 10.1103/PhysRevC.103.054909} {\bibfield  {journal} {\bibinfo  {journal} {Phys. Rev. C}\ }\textbf {\bibinfo {volume} {103}},\ \bibinfo {pages} {054909} (\bibinfo {year} {2021})},\ \Eprint {http://arxiv.org/abs/2010.15134} {arXiv:2010.15134 [nucl-th]} \BibitemShut {NoStop}%
\bibitem [{\citenamefont {Parkkila}\ \emph {et~al.}(2021)\citenamefont {Parkkila}, \citenamefont {Onnerstad},\ and\ \citenamefont {Kim}}]{Parkkila:2021tqq}%
  \BibitemOpen
  \bibfield  {author} {\bibinfo {author} {\bibfnamefont {J.~E.}\ \bibnamefont {Parkkila}}, \bibinfo {author} {\bibfnamefont {A.}~\bibnamefont {Onnerstad}}, \ and\ \bibinfo {author} {\bibfnamefont {D.~J.}\ \bibnamefont {Kim}},\ }\href {\doibase 10.1103/PhysRevC.104.054904} {\bibfield  {journal} {\bibinfo  {journal} {Phys. Rev. C}\ }\textbf {\bibinfo {volume} {104}},\ \bibinfo {pages} {054904} (\bibinfo {year} {2021})},\ \Eprint {http://arxiv.org/abs/2106.05019} {arXiv:2106.05019 [hep-ph]} \BibitemShut {NoStop}%
\bibitem [{\citenamefont {Arnold}\ \emph {et~al.}(2006)\citenamefont {Arnold}, \citenamefont {Dogan},\ and\ \citenamefont {Moore}}]{Arnold:2006fz}%
  \BibitemOpen
  \bibfield  {author} {\bibinfo {author} {\bibfnamefont {P.~B.}\ \bibnamefont {Arnold}}, \bibinfo {author} {\bibfnamefont {C.}~\bibnamefont {Dogan}}, \ and\ \bibinfo {author} {\bibfnamefont {G.~D.}\ \bibnamefont {Moore}},\ }\href {\doibase 10.1103/PhysRevD.74.085021} {\bibfield  {journal} {\bibinfo  {journal} {Phys. Rev. D}\ }\textbf {\bibinfo {volume} {74}},\ \bibinfo {pages} {085021} (\bibinfo {year} {2006})},\ \Eprint {http://arxiv.org/abs/hep-ph/0608012} {arXiv:hep-ph/0608012} \BibitemShut {NoStop}%
\bibitem [{\citenamefont {Arnold}\ \emph {et~al.}(2000)\citenamefont {Arnold}, \citenamefont {Moore},\ and\ \citenamefont {Yaffe}}]{Arnold:2000dr}%
  \BibitemOpen
  \bibfield  {author} {\bibinfo {author} {\bibfnamefont {P.~B.}\ \bibnamefont {Arnold}}, \bibinfo {author} {\bibfnamefont {G.~D.}\ \bibnamefont {Moore}}, \ and\ \bibinfo {author} {\bibfnamefont {L.~G.}\ \bibnamefont {Yaffe}},\ }\href {\doibase 10.1088/1126-6708/2000/11/001} {\bibfield  {journal} {\bibinfo  {journal} {JHEP}\ }\textbf {\bibinfo {volume} {11}},\ \bibinfo {pages} {001} (\bibinfo {year} {2000})},\ \Eprint {http://arxiv.org/abs/hep-ph/0010177} {arXiv:hep-ph/0010177} \BibitemShut {NoStop}%
\bibitem [{\citenamefont {Arnold}\ \emph {et~al.}(2003)\citenamefont {Arnold}, \citenamefont {Moore},\ and\ \citenamefont {Yaffe}}]{Arnold:2003zc}%
  \BibitemOpen
  \bibfield  {author} {\bibinfo {author} {\bibfnamefont {P.~B.}\ \bibnamefont {Arnold}}, \bibinfo {author} {\bibfnamefont {G.~D.}\ \bibnamefont {Moore}}, \ and\ \bibinfo {author} {\bibfnamefont {L.~G.}\ \bibnamefont {Yaffe}},\ }\href {\doibase 10.1088/1126-6708/2003/05/051} {\bibfield  {journal} {\bibinfo  {journal} {JHEP}\ }\textbf {\bibinfo {volume} {05}},\ \bibinfo {pages} {051} (\bibinfo {year} {2003})},\ \Eprint {http://arxiv.org/abs/hep-ph/0302165} {arXiv:hep-ph/0302165} \BibitemShut {NoStop}%
\bibitem [{\citenamefont {Sch\"afer}\ and\ \citenamefont {Teaney}(2009)}]{Schafer:2009dj}%
  \BibitemOpen
  \bibfield  {author} {\bibinfo {author} {\bibfnamefont {T.}~\bibnamefont {Sch\"afer}}\ and\ \bibinfo {author} {\bibfnamefont {D.}~\bibnamefont {Teaney}},\ }\href {\doibase 10.1088/0034-4885/72/12/126001} {\bibfield  {journal} {\bibinfo  {journal} {Rept. Prog. Phys.}\ }\textbf {\bibinfo {volume} {72}},\ \bibinfo {pages} {126001} (\bibinfo {year} {2009})},\ \Eprint {http://arxiv.org/abs/0904.3107} {arXiv:0904.3107 [hep-ph]} \BibitemShut {NoStop}%
\bibitem [{\citenamefont {Jeon}\ and\ \citenamefont {Yaffe}(1996)}]{Jeon:1995zm}%
  \BibitemOpen
  \bibfield  {author} {\bibinfo {author} {\bibfnamefont {S.}~\bibnamefont {Jeon}}\ and\ \bibinfo {author} {\bibfnamefont {L.~G.}\ \bibnamefont {Yaffe}},\ }\href {\doibase 10.1103/PhysRevD.53.5799} {\bibfield  {journal} {\bibinfo  {journal} {Phys. Rev. D}\ }\textbf {\bibinfo {volume} {53}},\ \bibinfo {pages} {5799} (\bibinfo {year} {1996})},\ \Eprint {http://arxiv.org/abs/hep-ph/9512263} {arXiv:hep-ph/9512263} \BibitemShut {NoStop}%
\bibitem [{\citenamefont {Sasaki}\ and\ \citenamefont {Redlich}(2009)}]{Sasaki:2008fg}%
  \BibitemOpen
  \bibfield  {author} {\bibinfo {author} {\bibfnamefont {C.}~\bibnamefont {Sasaki}}\ and\ \bibinfo {author} {\bibfnamefont {K.}~\bibnamefont {Redlich}},\ }\href {\doibase 10.1103/PhysRevC.79.055207} {\bibfield  {journal} {\bibinfo  {journal} {Phys. Rev. C}\ }\textbf {\bibinfo {volume} {79}},\ \bibinfo {pages} {055207} (\bibinfo {year} {2009})},\ \Eprint {http://arxiv.org/abs/0806.4745} {arXiv:0806.4745 [hep-ph]} \BibitemShut {NoStop}%
\bibitem [{\citenamefont {Bluhm}\ \emph {et~al.}(2007)\citenamefont {Bluhm}, \citenamefont {Kampfer}, \citenamefont {Schulze}, \citenamefont {Seipt},\ and\ \citenamefont {Heinz}}]{Bluhm:2007nu}%
  \BibitemOpen
  \bibfield  {author} {\bibinfo {author} {\bibfnamefont {M.}~\bibnamefont {Bluhm}}, \bibinfo {author} {\bibfnamefont {B.}~\bibnamefont {Kampfer}}, \bibinfo {author} {\bibfnamefont {R.}~\bibnamefont {Schulze}}, \bibinfo {author} {\bibfnamefont {D.}~\bibnamefont {Seipt}}, \ and\ \bibinfo {author} {\bibfnamefont {U.}~\bibnamefont {Heinz}},\ }\href {\doibase 10.1103/PhysRevC.76.034901} {\bibfield  {journal} {\bibinfo  {journal} {Phys. Rev. C}\ }\textbf {\bibinfo {volume} {76}},\ \bibinfo {pages} {034901} (\bibinfo {year} {2007})},\ \Eprint {http://arxiv.org/abs/0705.0397} {arXiv:0705.0397 [hep-ph]} \BibitemShut {NoStop}%
\bibitem [{\citenamefont {Bluhm}\ \emph {et~al.}(2012)\citenamefont {Bluhm}, \citenamefont {Kampfer},\ and\ \citenamefont {Redlich}}]{Bluhm:2011xu}%
  \BibitemOpen
  \bibfield  {author} {\bibinfo {author} {\bibfnamefont {M.}~\bibnamefont {Bluhm}}, \bibinfo {author} {\bibfnamefont {B.}~\bibnamefont {Kampfer}}, \ and\ \bibinfo {author} {\bibfnamefont {K.}~\bibnamefont {Redlich}},\ }\href {\doibase 10.1016/j.physletb.2012.01.069} {\bibfield  {journal} {\bibinfo  {journal} {Phys. Lett. B}\ }\textbf {\bibinfo {volume} {709}},\ \bibinfo {pages} {77} (\bibinfo {year} {2012})},\ \Eprint {http://arxiv.org/abs/1101.3072} {arXiv:1101.3072 [hep-ph]} \BibitemShut {NoStop}%
\bibitem [{\citenamefont {Chakraborty}\ and\ \citenamefont {Kapusta}(2011)}]{Chakraborty:2010fr}%
  \BibitemOpen
  \bibfield  {author} {\bibinfo {author} {\bibfnamefont {P.}~\bibnamefont {Chakraborty}}\ and\ \bibinfo {author} {\bibfnamefont {J.~I.}\ \bibnamefont {Kapusta}},\ }\href {\doibase 10.1103/PhysRevC.83.014906} {\bibfield  {journal} {\bibinfo  {journal} {Phys. Rev. C}\ }\textbf {\bibinfo {volume} {83}},\ \bibinfo {pages} {014906} (\bibinfo {year} {2011})},\ \Eprint {http://arxiv.org/abs/1006.0257} {arXiv:1006.0257 [nucl-th]} \BibitemShut {NoStop}%
\bibitem [{\citenamefont {Plumari}\ \emph {et~al.}(2011)\citenamefont {Plumari}, \citenamefont {Alberico}, \citenamefont {Greco},\ and\ \citenamefont {Ratti}}]{Plumari:2011mk}%
  \BibitemOpen
  \bibfield  {author} {\bibinfo {author} {\bibfnamefont {S.}~\bibnamefont {Plumari}}, \bibinfo {author} {\bibfnamefont {W.~M.}\ \bibnamefont {Alberico}}, \bibinfo {author} {\bibfnamefont {V.}~\bibnamefont {Greco}}, \ and\ \bibinfo {author} {\bibfnamefont {C.}~\bibnamefont {Ratti}},\ }\href {\doibase 10.1103/PhysRevD.84.094004} {\bibfield  {journal} {\bibinfo  {journal} {Phys. Rev. D}\ }\textbf {\bibinfo {volume} {84}},\ \bibinfo {pages} {094004} (\bibinfo {year} {2011})},\ \Eprint {http://arxiv.org/abs/1103.5611} {arXiv:1103.5611 [hep-ph]} \BibitemShut {NoStop}%
\bibitem [{\citenamefont {Alqahtani}\ \emph {et~al.}(2015)\citenamefont {Alqahtani}, \citenamefont {Nopoush},\ and\ \citenamefont {Strickland}}]{Alqahtani:2015qja}%
  \BibitemOpen
  \bibfield  {author} {\bibinfo {author} {\bibfnamefont {M.}~\bibnamefont {Alqahtani}}, \bibinfo {author} {\bibfnamefont {M.}~\bibnamefont {Nopoush}}, \ and\ \bibinfo {author} {\bibfnamefont {M.}~\bibnamefont {Strickland}},\ }\href {\doibase 10.1103/PhysRevC.92.054910} {\bibfield  {journal} {\bibinfo  {journal} {Phys. Rev. C}\ }\textbf {\bibinfo {volume} {92}},\ \bibinfo {pages} {054910} (\bibinfo {year} {2015})},\ \Eprint {http://arxiv.org/abs/1509.02913} {arXiv:1509.02913 [hep-ph]} \BibitemShut {NoStop}%
\bibitem [{\citenamefont {Berrehrah}\ \emph {et~al.}(2016)\citenamefont {Berrehrah}, \citenamefont {Bratkovskaya}, \citenamefont {Steinert},\ and\ \citenamefont {Cassing}}]{Berrehrah:2016vzw}%
  \BibitemOpen
  \bibfield  {author} {\bibinfo {author} {\bibfnamefont {H.}~\bibnamefont {Berrehrah}}, \bibinfo {author} {\bibfnamefont {E.}~\bibnamefont {Bratkovskaya}}, \bibinfo {author} {\bibfnamefont {T.}~\bibnamefont {Steinert}}, \ and\ \bibinfo {author} {\bibfnamefont {W.}~\bibnamefont {Cassing}},\ }\href {\doibase 10.1142/S0218301316420039} {\bibfield  {journal} {\bibinfo  {journal} {Int. J. Mod. Phys. E}\ }\textbf {\bibinfo {volume} {25}},\ \bibinfo {pages} {1642003} (\bibinfo {year} {2016})},\ \Eprint {http://arxiv.org/abs/1605.02371} {arXiv:1605.02371 [hep-ph]} \BibitemShut {NoStop}%
\bibitem [{\citenamefont {Ozvenchuk}\ \emph {et~al.}(2013)\citenamefont {Ozvenchuk}, \citenamefont {Linnyk}, \citenamefont {Gorenstein}, \citenamefont {Bratkovskaya},\ and\ \citenamefont {Cassing}}]{Ozvenchuk:2012kh}%
  \BibitemOpen
  \bibfield  {author} {\bibinfo {author} {\bibfnamefont {V.}~\bibnamefont {Ozvenchuk}}, \bibinfo {author} {\bibfnamefont {O.}~\bibnamefont {Linnyk}}, \bibinfo {author} {\bibfnamefont {M.~I.}\ \bibnamefont {Gorenstein}}, \bibinfo {author} {\bibfnamefont {E.~L.}\ \bibnamefont {Bratkovskaya}}, \ and\ \bibinfo {author} {\bibfnamefont {W.}~\bibnamefont {Cassing}},\ }\href {\doibase 10.1103/PhysRevC.87.064903} {\bibfield  {journal} {\bibinfo  {journal} {Phys. Rev. C}\ }\textbf {\bibinfo {volume} {87}},\ \bibinfo {pages} {064903} (\bibinfo {year} {2013})},\ \Eprint {http://arxiv.org/abs/1212.5393} {arXiv:1212.5393 [hep-ph]} \BibitemShut {NoStop}%
\bibitem [{\citenamefont {Soloveva}\ \emph {et~al.}(2020)\citenamefont {Soloveva}, \citenamefont {Moreau},\ and\ \citenamefont {Bratkovskaya}}]{Soloveva:2019xph}%
  \BibitemOpen
  \bibfield  {author} {\bibinfo {author} {\bibfnamefont {O.}~\bibnamefont {Soloveva}}, \bibinfo {author} {\bibfnamefont {P.}~\bibnamefont {Moreau}}, \ and\ \bibinfo {author} {\bibfnamefont {E.}~\bibnamefont {Bratkovskaya}},\ }\href {\doibase 10.1103/PhysRevC.101.045203} {\bibfield  {journal} {\bibinfo  {journal} {Phys. Rev. C}\ }\textbf {\bibinfo {volume} {101}},\ \bibinfo {pages} {045203} (\bibinfo {year} {2020})},\ \Eprint {http://arxiv.org/abs/1911.08547} {arXiv:1911.08547 [nucl-th]} \BibitemShut {NoStop}%
\bibitem [{\citenamefont {Marty}\ \emph {et~al.}(2013)\citenamefont {Marty}, \citenamefont {Bratkovskaya}, \citenamefont {Cassing}, \citenamefont {Aichelin},\ and\ \citenamefont {Berrehrah}}]{Marty:2013ita}%
  \BibitemOpen
  \bibfield  {author} {\bibinfo {author} {\bibfnamefont {R.}~\bibnamefont {Marty}}, \bibinfo {author} {\bibfnamefont {E.}~\bibnamefont {Bratkovskaya}}, \bibinfo {author} {\bibfnamefont {W.}~\bibnamefont {Cassing}}, \bibinfo {author} {\bibfnamefont {J.}~\bibnamefont {Aichelin}}, \ and\ \bibinfo {author} {\bibfnamefont {H.}~\bibnamefont {Berrehrah}},\ }\href {\doibase 10.1103/PhysRevC.88.045204} {\bibfield  {journal} {\bibinfo  {journal} {Phys. Rev. C}\ }\textbf {\bibinfo {volume} {88}},\ \bibinfo {pages} {045204} (\bibinfo {year} {2013})},\ \Eprint {http://arxiv.org/abs/1305.7180} {arXiv:1305.7180 [hep-ph]} \BibitemShut {NoStop}%
\bibitem [{\citenamefont {Lang}\ and\ \citenamefont {Weise}(2014)}]{Lang:2013lla}%
  \BibitemOpen
  \bibfield  {author} {\bibinfo {author} {\bibfnamefont {R.}~\bibnamefont {Lang}}\ and\ \bibinfo {author} {\bibfnamefont {W.}~\bibnamefont {Weise}},\ }\href {\doibase 10.1140/epja/i2014-14063-4} {\bibfield  {journal} {\bibinfo  {journal} {Eur. Phys. J. A}\ }\textbf {\bibinfo {volume} {50}},\ \bibinfo {pages} {63} (\bibinfo {year} {2014})},\ \Eprint {http://arxiv.org/abs/1311.4628} {arXiv:1311.4628 [hep-ph]} \BibitemShut {NoStop}%
\bibitem [{\citenamefont {Czajka}\ and\ \citenamefont {Jeon}(2017)}]{Czajka:2017bod}%
  \BibitemOpen
  \bibfield  {author} {\bibinfo {author} {\bibfnamefont {A.}~\bibnamefont {Czajka}}\ and\ \bibinfo {author} {\bibfnamefont {S.}~\bibnamefont {Jeon}},\ }\href {\doibase 10.1103/PhysRevC.95.064906} {\bibfield  {journal} {\bibinfo  {journal} {Phys. Rev. C}\ }\textbf {\bibinfo {volume} {95}},\ \bibinfo {pages} {064906} (\bibinfo {year} {2017})},\ \Eprint {http://arxiv.org/abs/1701.07580} {arXiv:1701.07580 [nucl-th]} \BibitemShut {NoStop}%
\bibitem [{\citenamefont {Deb}\ \emph {et~al.}(2016)\citenamefont {Deb}, \citenamefont {Kadam},\ and\ \citenamefont {Mishra}}]{Deb:2016myz}%
  \BibitemOpen
  \bibfield  {author} {\bibinfo {author} {\bibfnamefont {P.}~\bibnamefont {Deb}}, \bibinfo {author} {\bibfnamefont {G.~P.}\ \bibnamefont {Kadam}}, \ and\ \bibinfo {author} {\bibfnamefont {H.}~\bibnamefont {Mishra}},\ }\href {\doibase 10.1103/PhysRevD.94.094002} {\bibfield  {journal} {\bibinfo  {journal} {Phys. Rev. D}\ }\textbf {\bibinfo {volume} {94}},\ \bibinfo {pages} {094002} (\bibinfo {year} {2016})},\ \Eprint {http://arxiv.org/abs/1603.01952} {arXiv:1603.01952 [hep-ph]} \BibitemShut {NoStop}%
\bibitem [{\citenamefont {Singha}\ \emph {et~al.}(2019)\citenamefont {Singha}, \citenamefont {Abhishek}, \citenamefont {Kadam}, \citenamefont {Ghosh},\ and\ \citenamefont {Mishra}}]{Singha:2017jmq}%
  \BibitemOpen
  \bibfield  {author} {\bibinfo {author} {\bibfnamefont {P.}~\bibnamefont {Singha}}, \bibinfo {author} {\bibfnamefont {A.}~\bibnamefont {Abhishek}}, \bibinfo {author} {\bibfnamefont {G.}~\bibnamefont {Kadam}}, \bibinfo {author} {\bibfnamefont {S.}~\bibnamefont {Ghosh}}, \ and\ \bibinfo {author} {\bibfnamefont {H.}~\bibnamefont {Mishra}},\ }\href {\doibase 10.1088/1361-6471/aaf256} {\bibfield  {journal} {\bibinfo  {journal} {J. Phys. G}\ }\textbf {\bibinfo {volume} {46}},\ \bibinfo {pages} {015201} (\bibinfo {year} {2019})},\ \Eprint {http://arxiv.org/abs/1705.03084} {arXiv:1705.03084 [nucl-th]} \BibitemShut {NoStop}%
\bibitem [{\citenamefont {Mykhaylova}\ \emph {et~al.}(2019)\citenamefont {Mykhaylova}, \citenamefont {Bluhm}, \citenamefont {Redlich},\ and\ \citenamefont {Sasaki}}]{Mykhaylova:2019wci}%
  \BibitemOpen
  \bibfield  {author} {\bibinfo {author} {\bibfnamefont {V.}~\bibnamefont {Mykhaylova}}, \bibinfo {author} {\bibfnamefont {M.}~\bibnamefont {Bluhm}}, \bibinfo {author} {\bibfnamefont {K.}~\bibnamefont {Redlich}}, \ and\ \bibinfo {author} {\bibfnamefont {C.}~\bibnamefont {Sasaki}},\ }\href {\doibase 10.1103/PhysRevD.100.034002} {\bibfield  {journal} {\bibinfo  {journal} {Phys. Rev. D}\ }\textbf {\bibinfo {volume} {100}},\ \bibinfo {pages} {034002} (\bibinfo {year} {2019})},\ \Eprint {http://arxiv.org/abs/1906.01697} {arXiv:1906.01697 [hep-ph]} \BibitemShut {NoStop}%
\bibitem [{\citenamefont {Mykhaylova}\ and\ \citenamefont {Sasaki}(2021)}]{Mykhaylova:2020pfk}%
  \BibitemOpen
  \bibfield  {author} {\bibinfo {author} {\bibfnamefont {V.}~\bibnamefont {Mykhaylova}}\ and\ \bibinfo {author} {\bibfnamefont {C.}~\bibnamefont {Sasaki}},\ }\href {\doibase 10.1103/PhysRevD.103.014007} {\bibfield  {journal} {\bibinfo  {journal} {Phys. Rev. D}\ }\textbf {\bibinfo {volume} {103}},\ \bibinfo {pages} {014007} (\bibinfo {year} {2021})},\ \Eprint {http://arxiv.org/abs/2007.06846} {arXiv:2007.06846 [hep-ph]} \BibitemShut {NoStop}%
\bibitem [{\citenamefont {Rocha}\ \emph {et~al.}(2021)\citenamefont {Rocha}, \citenamefont {Denicol},\ and\ \citenamefont {Noronha}}]{Rocha:2021zcw}%
  \BibitemOpen
  \bibfield  {author} {\bibinfo {author} {\bibfnamefont {G.~S.}\ \bibnamefont {Rocha}}, \bibinfo {author} {\bibfnamefont {G.~S.}\ \bibnamefont {Denicol}}, \ and\ \bibinfo {author} {\bibfnamefont {J.}~\bibnamefont {Noronha}},\ }\href {\doibase 10.1103/PhysRevLett.127.042301} {\bibfield  {journal} {\bibinfo  {journal} {Phys. Rev. Lett.}\ }\textbf {\bibinfo {volume} {127}},\ \bibinfo {pages} {042301} (\bibinfo {year} {2021})},\ \Eprint {http://arxiv.org/abs/2103.07489} {arXiv:2103.07489 [nucl-th]} \BibitemShut {NoStop}%
\bibitem [{\citenamefont {Dash}\ \emph {et~al.}(2022)\citenamefont {Dash}, \citenamefont {Bhadury}, \citenamefont {Jaiswal},\ and\ \citenamefont {Jaiswal}}]{Dash:2021ibx}%
  \BibitemOpen
  \bibfield  {author} {\bibinfo {author} {\bibfnamefont {D.}~\bibnamefont {Dash}}, \bibinfo {author} {\bibfnamefont {S.}~\bibnamefont {Bhadury}}, \bibinfo {author} {\bibfnamefont {S.}~\bibnamefont {Jaiswal}}, \ and\ \bibinfo {author} {\bibfnamefont {A.}~\bibnamefont {Jaiswal}},\ }\href {\doibase 10.1016/j.physletb.2022.137202} {\bibfield  {journal} {\bibinfo  {journal} {Phys. Lett. B}\ }\textbf {\bibinfo {volume} {831}},\ \bibinfo {pages} {137202} (\bibinfo {year} {2022})},\ \Eprint {http://arxiv.org/abs/2112.14581} {arXiv:2112.14581 [nucl-th]} \BibitemShut {NoStop}%
\bibitem [{\citenamefont {Anderson}\ and\ \citenamefont {Witting}(1974)}]{anderson1974relativistic}%
  \BibitemOpen
  \bibfield  {author} {\bibinfo {author} {\bibfnamefont {J.~L.}\ \bibnamefont {Anderson}}\ and\ \bibinfo {author} {\bibfnamefont {H.}~\bibnamefont {Witting}},\ }\href@noop {} {\bibfield  {journal} {\bibinfo  {journal} {Physica}\ }\textbf {\bibinfo {volume} {74}},\ \bibinfo {pages} {466} (\bibinfo {year} {1974})}\BibitemShut {NoStop}%
\bibitem [{\citenamefont {Dash}\ and\ \citenamefont {Roy}(2020)}]{Dash:2020zqx}%
  \BibitemOpen
  \bibfield  {author} {\bibinfo {author} {\bibfnamefont {A.}~\bibnamefont {Dash}}\ and\ \bibinfo {author} {\bibfnamefont {V.}~\bibnamefont {Roy}},\ }\href {\doibase 10.1016/j.physletb.2020.135481} {\bibfield  {journal} {\bibinfo  {journal} {Phys. Lett. B}\ }\textbf {\bibinfo {volume} {806}},\ \bibinfo {pages} {135481} (\bibinfo {year} {2020})},\ \Eprint {http://arxiv.org/abs/2001.10756} {arXiv:2001.10756 [nucl-th]} \BibitemShut {NoStop}%
\bibitem [{\citenamefont {Biswas}\ \emph {et~al.}(2022)\citenamefont {Biswas}, \citenamefont {Mitra},\ and\ \citenamefont {Roy}}]{Biswas:2022cla}%
  \BibitemOpen
  \bibfield  {author} {\bibinfo {author} {\bibfnamefont {R.}~\bibnamefont {Biswas}}, \bibinfo {author} {\bibfnamefont {S.}~\bibnamefont {Mitra}}, \ and\ \bibinfo {author} {\bibfnamefont {V.}~\bibnamefont {Roy}},\ }\href {\doibase 10.1103/PhysRevD.106.L011501} {\bibfield  {journal} {\bibinfo  {journal} {Phys. Rev. D}\ }\textbf {\bibinfo {volume} {106}},\ \bibinfo {pages} {L011501} (\bibinfo {year} {2022})},\ \Eprint {http://arxiv.org/abs/2202.08685} {arXiv:2202.08685 [nucl-th]} \BibitemShut {NoStop}%
\bibitem [{\citenamefont {Rocha}\ \emph {et~al.}(2022)\citenamefont {Rocha}, \citenamefont {Ferreira}, \citenamefont {Denicol},\ and\ \citenamefont {Noronha}}]{Rocha:2022fqz}%
  \BibitemOpen
  \bibfield  {author} {\bibinfo {author} {\bibfnamefont {G.~S.}\ \bibnamefont {Rocha}}, \bibinfo {author} {\bibfnamefont {M.~N.}\ \bibnamefont {Ferreira}}, \bibinfo {author} {\bibfnamefont {G.~S.}\ \bibnamefont {Denicol}}, \ and\ \bibinfo {author} {\bibfnamefont {J.}~\bibnamefont {Noronha}},\ }\href {\doibase 10.1103/PhysRevD.106.036022} {\bibfield  {journal} {\bibinfo  {journal} {Phys. Rev. D}\ }\textbf {\bibinfo {volume} {106}},\ \bibinfo {pages} {036022} (\bibinfo {year} {2022})},\ \Eprint {http://arxiv.org/abs/2203.15571} {arXiv:2203.15571 [nucl-th]} \BibitemShut {NoStop}%
\bibitem [{\citenamefont {Gorenstein}\ and\ \citenamefont {Yang}(1995)}]{Gorenstein:1995vm}%
  \BibitemOpen
  \bibfield  {author} {\bibinfo {author} {\bibfnamefont {M.~I.}\ \bibnamefont {Gorenstein}}\ and\ \bibinfo {author} {\bibfnamefont {S.-N.}\ \bibnamefont {Yang}},\ }\href {\doibase 10.1103/PhysRevD.52.5206} {\bibfield  {journal} {\bibinfo  {journal} {Phys. Rev. D}\ }\textbf {\bibinfo {volume} {52}},\ \bibinfo {pages} {5206} (\bibinfo {year} {1995})}\BibitemShut {NoStop}%
\bibitem [{\citenamefont {Jeon}(1995)}]{Jeon:1994if}%
  \BibitemOpen
  \bibfield  {author} {\bibinfo {author} {\bibfnamefont {S.}~\bibnamefont {Jeon}},\ }\href {\doibase 10.1103/PhysRevD.52.3591} {\bibfield  {journal} {\bibinfo  {journal} {Phys. Rev. D}\ }\textbf {\bibinfo {volume} {52}},\ \bibinfo {pages} {3591} (\bibinfo {year} {1995})},\ \Eprint {http://arxiv.org/abs/hep-ph/9409250} {arXiv:hep-ph/9409250} \BibitemShut {NoStop}%
\bibitem [{\citenamefont {Romatschke}(2012)}]{Romatschke:2011qp}%
  \BibitemOpen
  \bibfield  {author} {\bibinfo {author} {\bibfnamefont {P.}~\bibnamefont {Romatschke}},\ }\href {\doibase 10.1103/PhysRevD.85.065012} {\bibfield  {journal} {\bibinfo  {journal} {Phys. Rev. D}\ }\textbf {\bibinfo {volume} {85}},\ \bibinfo {pages} {065012} (\bibinfo {year} {2012})},\ \Eprint {http://arxiv.org/abs/1108.5561} {arXiv:1108.5561 [gr-qc]} \BibitemShut {NoStop}%
\bibitem [{\citenamefont {Albright}\ and\ \citenamefont {Kapusta}(2016)}]{Albright:2015fpa}%
  \BibitemOpen
  \bibfield  {author} {\bibinfo {author} {\bibfnamefont {M.}~\bibnamefont {Albright}}\ and\ \bibinfo {author} {\bibfnamefont {J.~I.}\ \bibnamefont {Kapusta}},\ }\href {\doibase 10.1103/PhysRevC.93.014903} {\bibfield  {journal} {\bibinfo  {journal} {Phys. Rev. C}\ }\textbf {\bibinfo {volume} {93}},\ \bibinfo {pages} {014903} (\bibinfo {year} {2016})},\ \Eprint {http://arxiv.org/abs/1508.02696} {arXiv:1508.02696 [nucl-th]} \BibitemShut {NoStop}%
\bibitem [{\citenamefont {Tinti}\ \emph {et~al.}(2017)\citenamefont {Tinti}, \citenamefont {Jaiswal},\ and\ \citenamefont {Ryblewski}}]{Tinti:2016bav}%
  \BibitemOpen
  \bibfield  {author} {\bibinfo {author} {\bibfnamefont {L.}~\bibnamefont {Tinti}}, \bibinfo {author} {\bibfnamefont {A.}~\bibnamefont {Jaiswal}}, \ and\ \bibinfo {author} {\bibfnamefont {R.}~\bibnamefont {Ryblewski}},\ }\href {\doibase 10.1103/PhysRevD.95.054007} {\bibfield  {journal} {\bibinfo  {journal} {Phys. Rev. D}\ }\textbf {\bibinfo {volume} {95}},\ \bibinfo {pages} {054007} (\bibinfo {year} {2017})},\ \Eprint {http://arxiv.org/abs/1612.07329} {arXiv:1612.07329 [nucl-th]} \BibitemShut {NoStop}%
\bibitem [{\citenamefont {Czajka}\ \emph {et~al.}(2018)\citenamefont {Czajka}, \citenamefont {Hauksson}, \citenamefont {Shen}, \citenamefont {Jeon},\ and\ \citenamefont {Gale}}]{Czajka:2017wdo}%
  \BibitemOpen
  \bibfield  {author} {\bibinfo {author} {\bibfnamefont {A.}~\bibnamefont {Czajka}}, \bibinfo {author} {\bibfnamefont {S.}~\bibnamefont {Hauksson}}, \bibinfo {author} {\bibfnamefont {C.}~\bibnamefont {Shen}}, \bibinfo {author} {\bibfnamefont {S.}~\bibnamefont {Jeon}}, \ and\ \bibinfo {author} {\bibfnamefont {C.}~\bibnamefont {Gale}},\ }\href {\doibase 10.1103/PhysRevC.97.044914} {\bibfield  {journal} {\bibinfo  {journal} {Phys. Rev. C}\ }\textbf {\bibinfo {volume} {97}},\ \bibinfo {pages} {044914} (\bibinfo {year} {2018})},\ \Eprint {http://arxiv.org/abs/1712.05905} {arXiv:1712.05905 [nucl-th]} \BibitemShut {NoStop}%
\bibitem [{\citenamefont {Strickland}(2018)}]{Strickland:2018ayk}%
  \BibitemOpen
  \bibfield  {author} {\bibinfo {author} {\bibfnamefont {M.}~\bibnamefont {Strickland}},\ }\href {\doibase 10.1007/JHEP12(2018)128} {\bibfield  {journal} {\bibinfo  {journal} {JHEP}\ }\textbf {\bibinfo {volume} {12}},\ \bibinfo {pages} {128} (\bibinfo {year} {2018})},\ \Eprint {http://arxiv.org/abs/1809.01200} {arXiv:1809.01200 [nucl-th]} \BibitemShut {NoStop}%
\bibitem [{\citenamefont {Jaiswal}\ \emph {et~al.}(2019)\citenamefont {Jaiswal}, \citenamefont {Chattopadhyay}, \citenamefont {Jaiswal}, \citenamefont {Pal},\ and\ \citenamefont {Heinz}}]{Jaiswal:2019cju}%
  \BibitemOpen
  \bibfield  {author} {\bibinfo {author} {\bibfnamefont {S.}~\bibnamefont {Jaiswal}}, \bibinfo {author} {\bibfnamefont {C.}~\bibnamefont {Chattopadhyay}}, \bibinfo {author} {\bibfnamefont {A.}~\bibnamefont {Jaiswal}}, \bibinfo {author} {\bibfnamefont {S.}~\bibnamefont {Pal}}, \ and\ \bibinfo {author} {\bibfnamefont {U.}~\bibnamefont {Heinz}},\ }\href {\doibase 10.1103/PhysRevC.100.034901} {\bibfield  {journal} {\bibinfo  {journal} {Phys. Rev. C}\ }\textbf {\bibinfo {volume} {100}},\ \bibinfo {pages} {034901} (\bibinfo {year} {2019})},\ \Eprint {http://arxiv.org/abs/1907.07965} {arXiv:1907.07965 [nucl-th]} \BibitemShut {NoStop}%
\bibitem [{\citenamefont {Chattopadhyay}\ \emph {et~al.}(2023)\citenamefont {Chattopadhyay}, \citenamefont {Heinz},\ and\ \citenamefont {Schaefer}}]{Chattopadhyay:2023hpd}%
  \BibitemOpen
  \bibfield  {author} {\bibinfo {author} {\bibfnamefont {C.}~\bibnamefont {Chattopadhyay}}, \bibinfo {author} {\bibfnamefont {U.}~\bibnamefont {Heinz}}, \ and\ \bibinfo {author} {\bibfnamefont {T.}~\bibnamefont {Schaefer}},\ }\href {\doibase 10.1103/PhysRevC.108.034907} {\bibfield  {journal} {\bibinfo  {journal} {Phys. Rev. C}\ }\textbf {\bibinfo {volume} {108}},\ \bibinfo {pages} {034907} (\bibinfo {year} {2023})},\ \Eprint {http://arxiv.org/abs/2307.10769} {arXiv:2307.10769 [hep-ph]} \BibitemShut {NoStop}%
\bibitem [{\citenamefont {Dusling}\ \emph {et~al.}(2010)\citenamefont {Dusling}, \citenamefont {Moore},\ and\ \citenamefont {Teaney}}]{Dusling:2009df}%
  \BibitemOpen
  \bibfield  {author} {\bibinfo {author} {\bibfnamefont {K.}~\bibnamefont {Dusling}}, \bibinfo {author} {\bibfnamefont {G.~D.}\ \bibnamefont {Moore}}, \ and\ \bibinfo {author} {\bibfnamefont {D.}~\bibnamefont {Teaney}},\ }\href {\doibase 10.1103/PhysRevC.81.034907} {\bibfield  {journal} {\bibinfo  {journal} {Phys. Rev. C}\ }\textbf {\bibinfo {volume} {81}},\ \bibinfo {pages} {034907} (\bibinfo {year} {2010})},\ \Eprint {http://arxiv.org/abs/0909.0754} {arXiv:0909.0754 [nucl-th]} \BibitemShut {NoStop}%
\bibitem [{\citenamefont {Denicol}\ \emph {et~al.}(2012)\citenamefont {Denicol}, \citenamefont {Niemi}, \citenamefont {Molnar},\ and\ \citenamefont {Rischke}}]{Denicol:2012cn}%
  \BibitemOpen
  \bibfield  {author} {\bibinfo {author} {\bibfnamefont {G.~S.}\ \bibnamefont {Denicol}}, \bibinfo {author} {\bibfnamefont {H.}~\bibnamefont {Niemi}}, \bibinfo {author} {\bibfnamefont {E.}~\bibnamefont {Molnar}}, \ and\ \bibinfo {author} {\bibfnamefont {D.~H.}\ \bibnamefont {Rischke}},\ }\href {\doibase 10.1103/PhysRevD.85.114047} {\bibfield  {journal} {\bibinfo  {journal} {Phys. Rev. D}\ }\textbf {\bibinfo {volume} {85}},\ \bibinfo {pages} {114047} (\bibinfo {year} {2012})},\ \bibinfo {note} {[Erratum: Phys.Rev.D 91, 039902 (2015)]},\ \Eprint {http://arxiv.org/abs/1202.4551} {arXiv:1202.4551 [nucl-th]} \BibitemShut {NoStop}%
\bibitem [{\citenamefont {De~Groot}(1980)}]{DeGroot:1980dk}%
  \BibitemOpen
  \bibfield  {author} {\bibinfo {author} {\bibfnamefont {S.~R.}\ \bibnamefont {De~Groot}},\ }\href@noop {} {\emph {\bibinfo {title} {{Relativistic Kinetic Theory. Principles and Applications}}}},\ edited by\ \bibinfo {editor} {\bibfnamefont {W.~A.}\ \bibnamefont {Van~Leeuwen}}\ and\ \bibinfo {editor} {\bibfnamefont {C.~G.}\ \bibnamefont {Van~Weert}}\ (\bibinfo {year} {1980})\BibitemShut {NoStop}%
\bibitem [{\citenamefont {Bhadury}\ \emph {et~al.}(2020)\citenamefont {Bhadury}, \citenamefont {Florkowski}, \citenamefont {Jaiswal},\ and\ \citenamefont {Ryblewski}}]{Bhadury:2020ngq}%
  \BibitemOpen
  \bibfield  {author} {\bibinfo {author} {\bibfnamefont {S.}~\bibnamefont {Bhadury}}, \bibinfo {author} {\bibfnamefont {W.}~\bibnamefont {Florkowski}}, \bibinfo {author} {\bibfnamefont {A.}~\bibnamefont {Jaiswal}}, \ and\ \bibinfo {author} {\bibfnamefont {R.}~\bibnamefont {Ryblewski}},\ }\href {\doibase 10.1103/PhysRevC.102.064910} {\bibfield  {journal} {\bibinfo  {journal} {Phys. Rev. C}\ }\textbf {\bibinfo {volume} {102}},\ \bibinfo {pages} {064910} (\bibinfo {year} {2020})},\ \Eprint {http://arxiv.org/abs/2006.14252} {arXiv:2006.14252 [hep-ph]} \BibitemShut {NoStop}%
\bibitem [{\citenamefont {Bhadury}(2025)}]{Bhadury:2024ckc}%
  \BibitemOpen
  \bibfield  {author} {\bibinfo {author} {\bibfnamefont {S.}~\bibnamefont {Bhadury}},\ }\href {\doibase 10.1103/PhysRevC.111.034909} {\bibfield  {journal} {\bibinfo  {journal} {Phys. Rev. C}\ }\textbf {\bibinfo {volume} {111}},\ \bibinfo {pages} {034909} (\bibinfo {year} {2025})},\ \Eprint {http://arxiv.org/abs/2408.14462} {arXiv:2408.14462 [hep-ph]} \BibitemShut {NoStop}%
\bibitem [{\citenamefont {Bellac}(2011)}]{Bellac:2011kqa}%
  \BibitemOpen
  \bibfield  {author} {\bibinfo {author} {\bibfnamefont {M.~L.}\ \bibnamefont {Bellac}},\ }\href {\doibase 10.1017/CBO9780511721700} {\emph {\bibinfo {title} {{Thermal Field Theory}}}},\ Cambridge Monographs on Mathematical Physics\ (\bibinfo  {publisher} {Cambridge University Press},\ \bibinfo {year} {2011})\BibitemShut {NoStop}%
\bibitem [{\citenamefont {Nopoush}\ \emph {et~al.}(2015)\citenamefont {Nopoush}, \citenamefont {Strickland}, \citenamefont {Ryblewski}, \citenamefont {Bazow}, \citenamefont {Heinz},\ and\ \citenamefont {Martinez}}]{Nopoush:2015yga}%
  \BibitemOpen
  \bibfield  {author} {\bibinfo {author} {\bibfnamefont {M.}~\bibnamefont {Nopoush}}, \bibinfo {author} {\bibfnamefont {M.}~\bibnamefont {Strickland}}, \bibinfo {author} {\bibfnamefont {R.}~\bibnamefont {Ryblewski}}, \bibinfo {author} {\bibfnamefont {D.}~\bibnamefont {Bazow}}, \bibinfo {author} {\bibfnamefont {U.}~\bibnamefont {Heinz}}, \ and\ \bibinfo {author} {\bibfnamefont {M.}~\bibnamefont {Martinez}},\ }\href {\doibase 10.1103/PhysRevC.92.044912} {\bibfield  {journal} {\bibinfo  {journal} {Phys. Rev. C}\ }\textbf {\bibinfo {volume} {92}},\ \bibinfo {pages} {044912} (\bibinfo {year} {2015})},\ \Eprint {http://arxiv.org/abs/1506.05278} {arXiv:1506.05278 [nucl-th]} \BibitemShut {NoStop}%
\bibitem [{\citenamefont {Borsanyi}\ \emph {et~al.}(2010)\citenamefont {Borsanyi}, \citenamefont {Endrodi}, \citenamefont {Fodor}, \citenamefont {Jakovac}, \citenamefont {Katz}, \citenamefont {Krieg}, \citenamefont {Ratti},\ and\ \citenamefont {Szabo}}]{Borsanyi:2010cj}%
  \BibitemOpen
  \bibfield  {author} {\bibinfo {author} {\bibfnamefont {S.}~\bibnamefont {Borsanyi}}, \bibinfo {author} {\bibfnamefont {G.}~\bibnamefont {Endrodi}}, \bibinfo {author} {\bibfnamefont {Z.}~\bibnamefont {Fodor}}, \bibinfo {author} {\bibfnamefont {A.}~\bibnamefont {Jakovac}}, \bibinfo {author} {\bibfnamefont {S.~D.}\ \bibnamefont {Katz}}, \bibinfo {author} {\bibfnamefont {S.}~\bibnamefont {Krieg}}, \bibinfo {author} {\bibfnamefont {C.}~\bibnamefont {Ratti}}, \ and\ \bibinfo {author} {\bibfnamefont {K.~K.}\ \bibnamefont {Szabo}},\ }\href {\doibase 10.1007/JHEP11(2010)077} {\bibfield  {journal} {\bibinfo  {journal} {JHEP}\ }\textbf {\bibinfo {volume} {11}},\ \bibinfo {pages} {077} (\bibinfo {year} {2010})},\ \Eprint {http://arxiv.org/abs/1007.2580} {arXiv:1007.2580 [hep-lat]} \BibitemShut {NoStop}%
\bibitem [{\citenamefont {Ma}\ \emph {et~al.}(2019)\citenamefont {Ma}, \citenamefont {Lin}, \citenamefont {Qian}, \citenamefont {Hama},\ and\ \citenamefont {Kodama}}]{Ma:2018bwf}%
  \BibitemOpen
  \bibfield  {author} {\bibinfo {author} {\bibfnamefont {H.-H.}\ \bibnamefont {Ma}}, \bibinfo {author} {\bibfnamefont {K.}~\bibnamefont {Lin}}, \bibinfo {author} {\bibfnamefont {W.-L.}\ \bibnamefont {Qian}}, \bibinfo {author} {\bibfnamefont {Y.}~\bibnamefont {Hama}}, \ and\ \bibinfo {author} {\bibfnamefont {T.}~\bibnamefont {Kodama}},\ }\href {\doibase 10.1103/PhysRevC.100.015206} {\bibfield  {journal} {\bibinfo  {journal} {Phys. Rev. C}\ }\textbf {\bibinfo {volume} {100}},\ \bibinfo {pages} {015206} (\bibinfo {year} {2019})},\ \Eprint {http://arxiv.org/abs/1804.05376} {arXiv:1804.05376 [hep-ph]} \BibitemShut {NoStop}%
\bibitem [{\citenamefont {Denicol}\ and\ \citenamefont {Noronha}(2024)}]{Denicol:2022bsq}%
  \BibitemOpen
  \bibfield  {author} {\bibinfo {author} {\bibfnamefont {G.~S.}\ \bibnamefont {Denicol}}\ and\ \bibinfo {author} {\bibfnamefont {J.}~\bibnamefont {Noronha}},\ }\href {\doibase 10.1016/j.physletb.2024.138487} {\bibfield  {journal} {\bibinfo  {journal} {Phys. Lett. B}\ }\textbf {\bibinfo {volume} {850}},\ \bibinfo {pages} {138487} (\bibinfo {year} {2024})},\ \Eprint {http://arxiv.org/abs/2209.10370} {arXiv:2209.10370 [nucl-th]} \BibitemShut {NoStop}%
\bibitem [{\citenamefont {Weinberg}(1971)}]{Weinberg:1971mx}%
  \BibitemOpen
  \bibfield  {author} {\bibinfo {author} {\bibfnamefont {S.}~\bibnamefont {Weinberg}},\ }\href {\doibase 10.1086/151073} {\bibfield  {journal} {\bibinfo  {journal} {Astrophys. J.}\ }\textbf {\bibinfo {volume} {168}},\ \bibinfo {pages} {175} (\bibinfo {year} {1971})}\BibitemShut {NoStop}%
\bibitem [{\citenamefont {Buchel}(2005)}]{Buchel:2005cv}%
  \BibitemOpen
  \bibfield  {author} {\bibinfo {author} {\bibfnamefont {A.}~\bibnamefont {Buchel}},\ }\href {\doibase 10.1103/PhysRevD.72.106002} {\bibfield  {journal} {\bibinfo  {journal} {Phys. Rev. D}\ }\textbf {\bibinfo {volume} {72}},\ \bibinfo {pages} {106002} (\bibinfo {year} {2005})},\ \Eprint {http://arxiv.org/abs/hep-th/0509083} {arXiv:hep-th/0509083} \BibitemShut {NoStop}%
\bibitem [{\citenamefont {Buchel}(2008)}]{Buchel:2007mf}%
  \BibitemOpen
  \bibfield  {author} {\bibinfo {author} {\bibfnamefont {A.}~\bibnamefont {Buchel}},\ }\href {\doibase 10.1016/j.physletb.2008.03.069} {\bibfield  {journal} {\bibinfo  {journal} {Phys. Lett. B}\ }\textbf {\bibinfo {volume} {663}},\ \bibinfo {pages} {286} (\bibinfo {year} {2008})},\ \Eprint {http://arxiv.org/abs/0708.3459} {arXiv:0708.3459 [hep-th]} \BibitemShut {NoStop}%
\bibitem [{\citenamefont {Daher}\ \emph {et~al.}(2024)\citenamefont {Daher}, \citenamefont {Tinti}, \citenamefont {Jaiswal},\ and\ \citenamefont {Ryblewski}}]{Daher:2024vxk}%
  \BibitemOpen
  \bibfield  {author} {\bibinfo {author} {\bibfnamefont {A.}~\bibnamefont {Daher}}, \bibinfo {author} {\bibfnamefont {L.}~\bibnamefont {Tinti}}, \bibinfo {author} {\bibfnamefont {A.}~\bibnamefont {Jaiswal}}, \ and\ \bibinfo {author} {\bibfnamefont {R.}~\bibnamefont {Ryblewski}},\ }\href@noop {} {\  (\bibinfo {year} {2024})},\ \Eprint {http://arxiv.org/abs/2412.06024} {arXiv:2412.06024 [hep-ph]} \BibitemShut {NoStop}%
\bibitem [{\citenamefont {Denicol}\ \emph {et~al.}(2018)\citenamefont {Denicol}, \citenamefont {Huang}, \citenamefont {Moln\'ar}, \citenamefont {Monteiro}, \citenamefont {Niemi}, \citenamefont {Noronha}, \citenamefont {Rischke},\ and\ \citenamefont {Wang}}]{Denicol:2018rbw}%
  \BibitemOpen
  \bibfield  {author} {\bibinfo {author} {\bibfnamefont {G.~S.}\ \bibnamefont {Denicol}}, \bibinfo {author} {\bibfnamefont {X.-G.}\ \bibnamefont {Huang}}, \bibinfo {author} {\bibfnamefont {E.}~\bibnamefont {Moln\'ar}}, \bibinfo {author} {\bibfnamefont {G.~M.}\ \bibnamefont {Monteiro}}, \bibinfo {author} {\bibfnamefont {H.}~\bibnamefont {Niemi}}, \bibinfo {author} {\bibfnamefont {J.}~\bibnamefont {Noronha}}, \bibinfo {author} {\bibfnamefont {D.~H.}\ \bibnamefont {Rischke}}, \ and\ \bibinfo {author} {\bibfnamefont {Q.}~\bibnamefont {Wang}},\ }\href {\doibase 10.1103/PhysRevD.98.076009} {\bibfield  {journal} {\bibinfo  {journal} {Phys. Rev. D}\ }\textbf {\bibinfo {volume} {98}},\ \bibinfo {pages} {076009} (\bibinfo {year} {2018})},\ \Eprint {http://arxiv.org/abs/1804.05210} {arXiv:1804.05210 [nucl-th]} \BibitemShut {NoStop}%
\bibitem [{\citenamefont {Panda}\ \emph {et~al.}(2021)\citenamefont {Panda}, \citenamefont {Dash}, \citenamefont {Biswas},\ and\ \citenamefont {Roy}}]{Panda:2020zhr}%
  \BibitemOpen
  \bibfield  {author} {\bibinfo {author} {\bibfnamefont {A.~K.}\ \bibnamefont {Panda}}, \bibinfo {author} {\bibfnamefont {A.}~\bibnamefont {Dash}}, \bibinfo {author} {\bibfnamefont {R.}~\bibnamefont {Biswas}}, \ and\ \bibinfo {author} {\bibfnamefont {V.}~\bibnamefont {Roy}},\ }\href {\doibase 10.1007/JHEP03(2021)216} {\bibfield  {journal} {\bibinfo  {journal} {JHEP}\ }\textbf {\bibinfo {volume} {03}},\ \bibinfo {pages} {216} (\bibinfo {year} {2021})},\ \Eprint {http://arxiv.org/abs/2011.01606} {arXiv:2011.01606 [nucl-th]} \BibitemShut {NoStop}%
\end{thebibliography}%

\end{document}